%% file: double_intervals_quantum_template.tex
\documentclass[a4paper,twocolumn,11pt,accepted=2025-03-18]{quantumarticle}
\pdfoutput=1

\usepackage{graphicx} 
\input{tikz_lib}
\newcommand{\be}{\begin{equation}}
	\newcommand{\ee}{\end{equation}}

\pgfplotsset{
	colormap={bright}{rgb255=(251,51,255) rgb255=(255,128,0) rgb255=(0,0,255)
		rgb255=(255,0,0) rgb255=(128,255,0) rgb255=(204,204,0) rgb255=(127,0,255)
		rgb255=(0,0,0)}
}
\pgfplotsset{
	colormap={lightbluepalette}{ rgb255=(200, 240, 250) rgb255=(135, 206, 250) rgb255=(115, 194, 251)
		rgb255=(124,158,217) rgb255=(96, 130, 182)}
}

\theoremstyle{plain} 
\theoremstyle{plain}        
\newtheorem{theorem}{Theorem}       

\newtheorem{lemma}{Lemma}
\newtheorem{definition}{Definition}

\definecolorseries{foo}{hsb}{step}[hsb]{0,1,1}[hsb]{.618,0,0}
\renewcommand{\vec}[1]{{\boldsymbol{#1}}}

\begin{document}
\title{Entanglement of Disjoint Intervals in Dual-Unitary Circuits: Exact Results}
	\author{Alessandro Foligno}
	\affiliation{School of Physics and Astronomy, University of Nottingham, Nottingham, NG7 2RD, UK}
	\affiliation{Centre for the Mathematics and Theoretical Physics of Quantum Non-Equilibrium Systems, University of Nottingham, Nottingham, NG7 2RD, UK}
	\author{Bruno Bertini}
	\affiliation{School of Physics and Astronomy, University of Birmingham, Edgbaston, Birmingham, B15 2TT, UK}
\begin{abstract}
The growth of the entanglement between two disjoint intervals and its complement after a quantum quench is regarded as a dynamical chaos indicator. Namely, it is expected to show qualitatively different behaviours depending on whether the underlying microscopic dynamics is chaotic or integrable. So far, however, this could only be verified in the context of conformal field theories. Here we present an exact confirmation of this expectation in a class of interacting microscopic Floquet systems on the lattice, i.e., dual-unitary circuits. These systems can either have \emph{zero} or a \emph{super extensive} number of conserved charges: the latter case is achieved via fine-tuning. We show that, for \emph{almost all} dual unitary circuits on qubits and for a large family of dual-unitary circuits on qudits the asymptotic entanglement dynamics agrees with what is expected for chaotic systems. On the other hand, if we require the systems to have conserved charges, we find that the entanglement displays the qualitatively different behaviour expected for integrable systems. Interestingly, despite having many conserved charges, charge-conserving dual-unitary circuits are in general not Yang-Baxter integrable.  
\end{abstract}
\maketitle

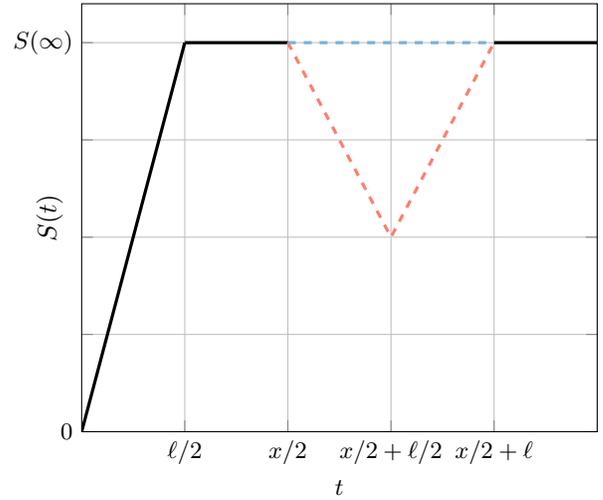
\begin{figure}[t]\comment{
	\begin{tikzpicture}[scale=1.0,remember picture]
	\begin{axis}[grid=major,colormap name=springpastels,
		cycle list={[of colormap]},
		legend columns=2,
		legend style={at={(0.6,0.7)},anchor=south east,font=\scriptsize,draw= none, fill=none},
		xtick distance=0.5,
		mark size=1.3pt,	
		xlabel=$t$,
		ylabel=$S(t)$,
		y label style={at={(axis description cs:0.16,.5)},anchor=south,font=\normalsize	},		
		tick label style={font=\small	},	
		x label style={font=\small	},	
		ytick={1,0.75,0.5,0.25,0},
		xtick={1,2,3,4},
		grid style={thin},
		xticklabels={${\ell}/{2}$,${x}/{2}$,$x/2+\ell/2$,${x}/{2}+\ell$},
		yticklabels={$S(\infty)$,$\phantom{S'}$,$\phantom{S'}$,$\phantom{S'}$,0},
		xmin=0,
		ymin=0,
		xmax=5
		]
		
		\addplot[domain=0:1,very thick,color=black]{x};
		\addplot[domain=1:2,very thick,color=black]{1};
		\addplot[domain=4:6,very thick,color=black]{1};
		\addplot[domain=2:4,very thick,dashed,color=pastelred]{0.5+0.5*abs(x-3)};
		\addplot[domain=2:4,very thick,dashed,color=pastelblue]{1};
	\end{axis}
\end{tikzpicture}}
\includegraphics{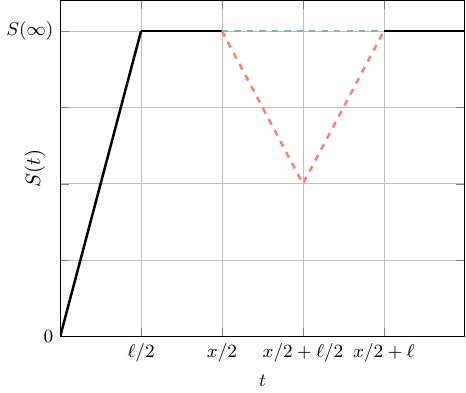}
\caption{The entanglement of a subsystem $A$ composed by two disjoint intervals of size $\ell$ separated by a distance $x>\ell$, i.e.\ $A=[0,\ell]\cup[\ell+x, 2\ell+x]$, according to the membrane picture (dashed blue) and the quasiparticle one (dashed red). The continuous black line refers to the points on which both predictions agree. The quasiparticles are taken to move at speed $v_{\rm qp}=1$.}
\label{fig:plot1}
\end{figure}

\section{Introduction}

Under very general conditions, a quantum quench triggers a linear growth of entanglement~\cite{calabrese2005evolution, fagotti2008evolution, alba2018entanglement, alba2017entanglement, alba2018entanglement,lagnese2021entanglement, liu2014entanglement, asplund2015entanglement, laeuchli2008spreading, kim2013ballistic, nahum2017quantum, nahum2018operator, vonKeyserlingk2018operator, pal2018entangling, bertini2019entanglement, piroli2020exact, gopalakrishnan2019unitary}. This phenomenon can be explained using a duality between space and time~\cite{bertini2022growth}, which interprets the linear growth as another manifestation of the extensivity of the stationary entropy. The underlying mechanisms driving the growth, however, are expected to depend on the nature of the dynamics, i.e., on whether the system is integrable or chaotic. In the former case, the growth is described in terms of the motion of correlated quasiparticles produced by the quench, while in the latter is connected with the expansion of the minimal space-time membrane separating the subsystems. These two mechanisms are phenomenologically described by the \emph{quasiparticle picture}~\cite{calabrese2005evolution} and the \emph{membrane picture}~\cite{jonay2018coarsegrained, zhou2020entanglement} respectively.

In  one dimension, both theories yield the same qualitative prediction --- linear growth followed by saturation --- for the entanglement of a single, connected interval\footnote{We use the terms ``entanglement'' and ``entanglement entropy'' interchangeably as the latter is a measure of bipartite entanglement for pure states, which is the case under investigation here.}. For subsystems with more complicated geometries, however, the predictions of the two theories differ, providing potential macroscopic manifestations of the nature of the microscopic dynamics. For instance, consider a subsystem $A$ made of two intervals of equal size $\ell$, separated by a distance $x\gg \ell$. After an initial linear growth phase followed by saturation, the quasiparticle picture predicts a temporary drop in the entanglement entropy when the two intervals become causally connected, i.e., for times $t\in [{x}/{2 v_{\rm qp}},{(x+\ell)}/{2 v_{\rm qp}}]$,
where $v_{\rm qp}$ is the quasiparticle speed\footnote{Note that in integrable models one generally has several species of quasiparticles with different velocities. This results in a smoothening of the curve in Fig.~\ref{fig:plot1}, see, e.g., Ref.~\cite{alba2019quantum}. Here, for simplicity, we focus on the case with a single species of quasiparticles.}. Instead, according to the membrane picture, once the entanglement saturates after the initial growth phase, its value remains constant, see Fig.~\ref{fig:plot1}.

Currently, however, the only exact results substantiating this picture have been obtained in conformal field theory~\cite{calabrese2009entanglement, asplund2015entanglement}, for non-interacting spin chains~\cite{fagotti2010entanglement}, and for  unitary circuits in the limit of large local Hilbert space dimension~\cite{nahum2017quantum}. No exact result exists for clean, microscopic systems in the presence of interactions. In this paper we fill this gap and present a rigorous proof of the occurrence of these different behaviours for  dual-unitary circuits~\cite{bertini2019exact}, a class of quantum circuits that has been extensively studied in recent years~\cite{bertini2019exact, prosen2021many, kos2021correlations, bertini2022exact, kos2023circuitsofspacetime,gopalakrishnan2019unitary,claeys2021ergodic}. These circuits generically exhibit chaotic behaviour, and we will indeed prove that they generically follow the membrane picture in Fig.~\ref{fig:plot1}, however, they can also be equipped with a charge structure~\cite{bertini2020operator,bertini2020operator2, holdendye2023fundamental}, which can drastically affect their entanglement dynamics~\cite{foligno2024nonequilibrium}. We will show that, although they are generically not Yang--Baxter-integrable, charged dual unitary circuits do follow the quasiparticle prediction in Fig.~\ref{fig:plot1}.

More specifically, the rest of the paper is laid out as follows. In Sec.~\ref{sec:setting} we describe setting and quench protocol under investigation. Then, in Sec.~\ref{sec:entanglementdisjoint}, we move on to the characterisation of the quantity of interest: the dynamics of the entanglement of two disjoint intervals after a quantum quench. In particular,  in Sec.~\ref{sec:entanglementchaotic} we consider the case of generic dual-unitary circuits while in Sec.~\ref{sec:entanglementcharged} we study charged ones. Finally, Sec.~\ref{sec:conclusions} contains our conclusions. Some of the technical derivations are reported in the three appendices.

\section{Setting}
\label{sec:setting}

\begin{figure*}[t]
	\comment{
\begin{tikzpicture}[scale=0.6]
{
		\def 	\talln {0}\pgfmathparse{2*(\talln+2)}
		\draw[fill=lightred,color=lightred] (2-2,-1)--++(0,4)--++(6,0)--++(0,-4)--cycle;
		\draw[fill=lightred,color=lightred] (12+2,-1)--++(0,4)--++(6,0)--++(0,-4)--cycle;
		\foreach \i in {0,...,\talln}
		{	
			\foreach \j in {-1,...,10}
			{
				\roundgate[2*\j][2*\i][1][topright][bertiniorange][-1]
				\roundgate[2*\j+1][2*\i+1][1][topright][bertiniorange][-1]
				\bellpair[2*\j+1][-1]
			}
		}
		\foreach \i in {-2,-1,6,7,8,9,10,11,12,13,20,21}
		{
			\cstate[\i+.5][1.5+2*\talln]
		}
		\pgfmathparse{2*(\talln+1)+.5}
		\node[scale=1.25] at (5-2,\pgfmathresult) {$A$};
		\node[scale=1.25] at (15+2,\pgfmathresult) {$A$};
		\node[scale=1.25] at (10,\pgfmathresult) {$2x$ qudits};
		\draw[decorate,decoration={brace}] (6.2,\pgfmathresult-.7)--++(7.6,0);
	}
\end{tikzpicture}
}
\includegraphics{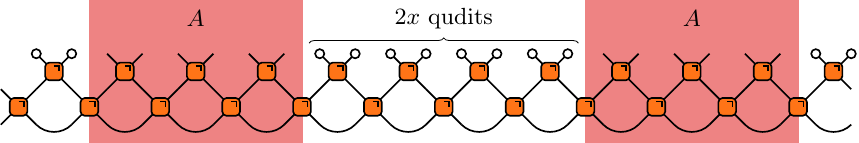}
\caption{Reduced density matrix of the subsystem $A=[0,\ell]\cup[\ell+x, 2\ell+x]$ according to the diagrammatic representation described in Eqs.~\eqref{eq:foldedgatepicture}--\eqref{eq:DUgraph}. The case depicted refers to the early-time scenario $t\le x$ when the two intervals are not causally connected, and their contribution to the entanglement factorises. We took $t=1$, $x=4$, and $\ell=3$.}
\label{fig:rhoA}
\end{figure*}

We consider locally interacting quantum many-body systems of qudits in discrete space time. After being prepared in some initial state $\ket{\Psi_0}$, the system evolves in discrete steps implemented by the many-body unitary operator $\mathbb{U}$, which is constructed in terms of a two-qudit gate $U$ applied in the following brickwork pattern  
\be
\mathbb{U} =\mathbb{U}_e\mathbb{U}_o,\,\, \mathbb{U}_e=\bigotimes_{x=1}^{L} U_{x,x+1/2},\,\,\mathbb{U}_o=\bigotimes_{x=0}^{L-1} U_{x+1/2,x}\,.
\label{eq:U}
\ee
Here the qudits sit at half integer positions, the operator $U_{a,b}$ acts as $U$ on the qudits located at $a$ and $b$, and we denote by $2 L$ the number of qudits, which we take to be much larger than all other quantities at play. Moreover, we consider periodic boundary conditions, so that sites $0$ and $L$ coincide, and indicate by $d$ the number of states of each qudit (local Hilbert space dimension). 

Specifically, here we focus on the case where the initial state is made of maximally entangled Bell pairs among nearest neighbours, i.e., 
\begin{align}
\label{eq:initstate}
	\ket{\Psi_0}=\bigotimes_{x=1}^L \ket{\psi_0}, \qquad \ket{\psi_0} = \sum_{i=1}^d \frac{\ket{i}_x\ket{i}_{x+1/2}}{\sqrt{d}},
\end{align}
and the local gate $U$ is \emph{dual-unitary} (DU)~\cite{bertini2019exact}, meaning that also the matrix $\widetilde{U}$ obtained as
\be
\bra{ij}\widetilde{U}\ket{kl}\equiv \mel{jl}{U}{ik},
\ee	
is a unitary matrix. Physically, this reshuffling corresponds to an exchange of space and time in the quantum circuit, therefore DU gates can be defined as those generating unitary dynamics in \emph{both} space \emph{and} time~\cite{bertini2019exact}. 

A complete parameterisation of DU gates is only known for $d=2$~\cite{bertini2019exact}, however, families of DU gates are known for any $d\geq2$~\cite{claeys2021ergodic, blockeddu,rather2020creating,roundaface,borsi2022construction}. Here we are interested in the following one  
\begin{align}
	\!\!\!\!\mathcal S_{\lt} = \{S \cdot U^{[u],\lt} \cdot\left(\mathds{1}_d	\otimes v \right) \quad v, u^{(1)}, \ldots, u^{(d)}\in U(d)\},\!\!\!\!\label{eq:DUl}
\end{align}
where $S$ is the SWAP gate and $U^{[u],\lt}$ is a control gate defined in terms of $d$ unitary matrices $\{u^{(i)}\}$ as 
\begin{align}
U^{[u],\lt}\left(\ket{i}\otimes\ket{j}\right)=u^{(j)}\ket{i}\otimes\ket{j}. 
\label{eq:controlright}
\end{align}
Here $\{\ket{j}\}$ denotes the computational basis. When seen as a manifold, the set $\mathcal S_{\lt}$ has a number of parameters scaling as $d^3$ for large $d$. This is the largest scaling observed for families of DU gates~\cite{roundaface,borsi2022construction}. The set $\mathcal S_{\lt}$, however, is not the only one with this property. For instance, one can construct a different family, $\mathcal S_{\rt}$, by applying a spatial reflection on the elements of $\mathcal S_{\lt}$. Namely, $\mathcal S_{\rt}=S \cdot \mathcal S_{\lt}\cdot S$. For $d=2$, both $\mathcal S_{\lt}$ and $\mathcal S_{\rt}$ coincide with the set of all DU gates~\cite{bertini2019exact}. To make the upcoming discussion more precise it is useful to introduce the following definition
\begin{definition}
\label{def:def1}
We call \emph{generic} a DU gate $U\in\mathcal S_{\lt/\rt}$ that is constructed with matrices $v, u^{(1)}, \ldots, u^{(d)}$ independently drawn from $U(d)$ according to its Haar measure. Similarly, we call \emph{generic} a DU circuit where the time evolution operator~\eqref{eq:U} is built with generic dual unitary gates. 
\end{definition}
We stress that there is no noise in the generic DU circuits introduced here: once the local gate is chosen, it is kept constant in both space and time.

The evolution of quantum circuits is conveniently represented graphically and here we follow this approach. Specifically, we introduce the following diagrammatic notation for gate and initial state (cf.~Eqs.~\eqref{eq:U} and \eqref{eq:initstate}), in a one-replica, or \emph{folded} space 
\be
U\otimes U^* =\fineq[-0.8ex][.8][1]{\roundgate[0][0][1][topright][bertiniorange][-1]}, \qquad \ketbra{\psi_0} =\frac{1}{d}	\fineq[-0.8ex][0.8][1]{\foreach \i in {0}{\bellpair[2*\i][0]}}. \label{eq:foldedgatepicture}
\ee
The local trace operation becomes a state which we represent with a white circle
\begin{align}
\ket{\mcirc}=\left(\sum_{i=1}^d \ket{i,i}\right)\equiv\fineq{\draw[thick] (0,0)--++(-.5,0);\cstate[0][0]}\,.
\label{eq:bullet}
\end{align}
In this notation, the dual-unitarity property is expressed by the graphical rules
\begin{align}
	\fineq[-0.8ex][.8][1]{
		\roundgate[1][1][1][topright][bertiniorange][-1]
		\cstate[1.5][1.5] 
		\cstate[.5][1.5] 
	}&=
	\fineq[-0.8ex][.8][1]{
		\draw (.5,1.5)--++(0,-1);
		\draw (1.5,1.5)--++(0,-1);
		\cstate[1.5][1.5] 
		\cstate[.5][1.5] },&
	\fineq[-0.8ex][.8][1]{
		\roundgate[1][1][1][topright][bertiniorange][-1]
		\cstate[1.5][.5] 
		\cstate[.5][.5] }&=\fineq[-0.8ex][.8][1]{
		\draw (.5,1.5)--++(0,-1);
		\draw (1.5,1.5)--++(0,-1);
		\cstate[.5][.5] 
		\cstate[1.5][.5] },\label{eq:unitarityfoldeddiagram}\\
	\fineq[-0.8ex][.8][1]{
		\roundgate[1][1][1][topright][bertiniorange][-1]
		\cstate[1.5][1.5] 
		\cstate[1.5][.5] 
	}&=
	\fineq[-0.8ex][.8][1]{
		\draw (1.5,1.5)--++(-1,0);
		\draw (1.5,.5)--++(-1,0);
		\cstate[1.5][1.5] 
		\cstate[1.5][.5] },&
	\fineq[-0.8ex][.8][1]{
		\roundgate[1][1][1][topright][bertiniorange][-1]		
		\cstate[.5][1.5] 
		\cstate[.5][.5] } &=\fineq[-0.8ex][.8][1]{
		\draw (1.5,.5)--++(-1,0);
		\draw (1.5,1.5)--++(-1,0);
		\cstate[.5][1.5] 
		\cstate[.5][.5] }.
	\label{eq:DUgraph}
\end{align}

\section{Entanglement of two disjoint intervals}
\label{sec:entanglementdisjoint}

In this paper we consider the evolution of entanglement between a region $A$ and its complement by measuring the entanglement entropy 
\be
S_A(t) \equiv - \tr[\rho_A(t)\log \rho_A(t)],
\ee
where $\rho_A(t)$ is the density matrix of $A$. As discussed in the introduction, our focus is on the case where $A$ is composed of two intervals of length $\ell$ (meaning they contain $2\ell$ qudits each in our units) separated by a region of length $x>\ell$. In this case, $\rho_A(t)$ is represented diagrammatically as in Fig.~\ref{fig:rhoA}. 

Whenever the two intervals remain causally disconnected, i.e.\ for $t\leq x/2$, the entanglement of $\rho_A(t)$ can be characterised exactly. Indeed, using the rules in Eq.~\ref{eq:DUgraph}, we can simplify the top diagram in Fig.~\ref{fig:rhoA} to  
\begin{align}
\rho_A(t) & \propto	\left(\fineq[-0.8ex][0.65][1]{
		\pgfmathparse{2.5}
		\roundgate[3][1][1][topright][bertiniorange][-1]						
		\roundgate[5][1][1][topright][bertiniorange][-1]				
		\roundgate[7][1][1][topright][bertiniorange][-1]	
		\roundgate[4][0][1][topright][bertiniorange][-1]						
		\roundgate[6][0][1][topright][bertiniorange][-1]					
		\bellpair[5][-1]
		\foreach \i in {0,1}
		{
			\cstate[+3.5-\i][\i-.5]
			\cstate[+6.5+\i][\i-.5]
		}
	}\right)^{\otimes 2}\notag\\
	& \simeq
	\left(\fineq[-0.8ex][0.55][1]
	{	
	\foreach \i in {0,1}
	{
	\draw[thick](4.-\i,4)--++(0,.5);
	\cstate[4.-\i][4]
	\draw[thick](7+\i,4)--++(0,.5);
	\cstate[7+\i][4]
	}
	\bellpair[5.5][4]
	}\right)^{\otimes 2},
\label{eq:rmdinitial2}
\end{align}
where in the second step we performed a similarity transformation within $A$, which does not change the entanglement, in order to remove the gates ($\ell=3$ in the above equation). This gives the following exact expression for entanglement entropy for $t\leq x/2$, which is the one obtained for a single interval in \cite{piroli2020exact} with an extra factor of $2$
 \begin{align}
	S_A(t) =4\min(2t,\ell) \log(d).
\end{align}
Applying Eqs.~\eqref{eq:unitarityfoldeddiagram}--\eqref{eq:DUgraph} one can also directly show that at times $t\ge {x}/{2}+{\ell}$ the state of the region $A$  relaxes to the infinite temperature state 
\be
\rho_A(t\ge {x}/{2}+{\ell})\propto \mathds{1}_{4\ell},
\ee
where $\mathds{1}_{x}$ represents the identity operator in $\mathbb C^{d^x}$. Therefore, we have 
\be
S_A(t\ge {x}/{2}+{\ell})=4\ell \log(d).
\ee

\begin{figure*}[t]
	\comment{
\begin{tikzpicture}[scale=0.6]
\foreach \i in {0,...,5}
		{
			\foreach \j in {-1,...,1}
			{
				\roundgate[-\i+\j][\i+\j][1][topright][bertiniorange][-1]
				\roundgate[4+\i-\j][\i+\j][1][topright][bertiniorange][-1]
			}
		}
		\draw[thick,dashed,red] (-6.5,2.5)--++(4,4)--++(1,-1)--++(-4,-4)--cycle;
		\node[scale=1.5,red] at (-1.8,6.8) {${T}^\lt_{z}$};
\begin{scope}[shift={(4,0)}]
		\draw[thick,dashed,red] (6.5,2.5)--++(-4,4)--++(-1,-1)--++(4,-4)--cycle;
\node[scale=1.5,red] at (1.8,6.8) {${T}^\rt_{z}$};
\end{scope}
		\roundgate[-6][6][1][topright][bertiniorange][-1]
		\roundgate[-8][6][1][topright][bertiniorange][-1]
		\roundgate[-7][5][1][topright][bertiniorange][-1]
		\roundgate[10][6][1][topright][bertiniorange][-1]
		\roundgate[12][6][1][topright][bertiniorange][-1]
		\roundgate[11][5][1][topright][bertiniorange][-1]
		\roundgate[1][-1][1][topright][bertiniorange][-1]
		\roundgate[3][-1][1][topright][bertiniorange][-1]
		\roundgate[2][0][1][topright][bertiniorange][-1]
		\foreach \i in {-1,...,6}
		{
			\cstate[-\i-2.5][\i-.5]
			\cstate[6+\i+.5][\i-.5]
		}
		\foreach \i in {0,...,4}
		{
			\cstate[-\i+1.5][\i+1.5]
			\cstate[4+\i-1.5][\i+1.5]
		}	
		\bellpair[2][-2]
		\bellpair[0][-2]
		\bellpair[4][-2]
		\draw[decorate,decoration={brace}] (2,1.5)--++(4.5,4.5);
		\node[scale=1.25] at (3.75,4.) {$x$};
\end{tikzpicture}}
\includegraphics{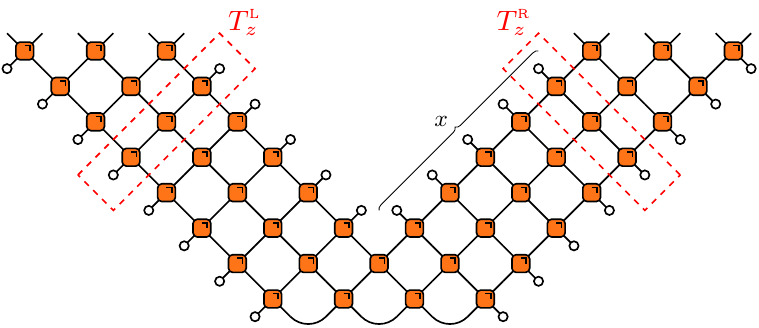}
\caption{Reduced density matrix of the subsystem $A=[0,\ell]\cup[\ell+x, 2\ell+x]$ according to the diagrammatic representation described in Eqs.~\eqref{eq:foldedgatepicture}--\eqref{eq:DUgraph}. The case depicted refers to the time regime where the two intervals are connected by some light-cones, i.e. at times $x/2<t<\ell+x/2$, which cannot be simplified using only dual unitarity. We took $t=4$, $x=5$, and $\ell=3$.}
\label{fig:rhoA2}
\end{figure*}

The only time window where the entanglement is not fixed by dual unitarity is thus $\smash{x/2<t<x/2+\ell}$. In this regime, the reduced density matrix can be simplified as the one depicted in Fig.~\ref{fig:rhoA2} and, as we show in the two upcoming subsections, shows very different behaviours depending on whether the circuit under investigation is generic (in the sense of Definition~\ref{def:def1}) or charged. 

\subsection{Generic Dual-Unitary Circuits}
\label{sec:entanglementchaotic}

Looking at the diagram in Fig.~\ref{fig:rhoA2} we see that it involves $x$ applications of the transfer matrices $T_{\ell}^{\lt/\rt}$ circled in red (the diagram in the figure has $\ell=3$). This implies that, for large $x$ (and fixed $\ell$), the diagram can be simplified by truncating the transfer matrices to their leading eigenspaces. For generic DU circuits, the leading eigenspaces of $T_{\ell}^\lt $ and $T_{\ell}^\rt $ are characterised by the following theorem, which is the first main result of this paper. 
\begin{theorem}
\label{thm:thm1}
Generic dual-unitary circuits in $\mathcal S_{\lt}$ produce \emph{almost surely} a matrix $T_{z}^\lt $ with a \emph{unique} maximal eigenvalue $d$ and right (left) eigenvector $\smash{\propto \ket{\mcirc}^{\otimes z}}$ ($\smash{\propto \bra{\mcirc}^{\otimes z}}$).  
\end{theorem}
\noindent An analogous result holds when replacing $\lt$ with $\rt$. \\

\begin{figure}
	\comment{
	\begin{tikzpicture}[scale=1,remember picture]
		\begin{axis}[grid=major,colormap name=springpastels,
			cycle list={[of colormap]},
			legend columns=1,
			legend style={at={(0.3,0.1)},anchor=south east,font=\scriptsize,draw= none, fill=none},
			xtick distance=0.2,
			mark size=1.3pt,	
			xlabel=$t/\ell$,
			ylabel=$ S(x/2+t)/(2\ell\log(2))$,
			y label style={at={(axis description cs:0.05,.5)},anchor=south,font=\normalsize	},		
			tick label style={font=\normalsize	},	
			x label style={font=\normalsize	},	
			xmax=1,
			xmin=0
			]
			
			
			\addplot[
			smooth,
			thick,
			mark=*,
			color=pastelgreen,
			] table[ x expr=((\thisrowno{0})	), y expr=(\thisrowno{1})] {plots_data/scaledentanglement_l=3.dat};
			\addlegendentry{$\ell=3$}
			\addlegendentry{$\ell=4$}
			\addlegendentry{$\ell=5$}
			\addlegendentry{$\ell=6$}
			
			\addplot[smooth,thick,mark=*,color=pastelred,] table[ x expr=((\thisrowno{0})	), y expr=(\thisrowno{1})] {plots_data/scaledentanglement_l=4.dat};
			
			

			\addplot[smooth,thick,mark=*,color=pastelblue,] table[ x expr=((\thisrowno{0})	), y expr=(\thisrowno{1})] {plots_data/scaledentanglement_l=5.dat};
			
			

			\addplot[smooth,thick,mark=*,color=newcol,] table[ x expr=((\thisrowno{0})	), y expr=(\thisrowno{1})] {plots_data/scaledentanglement_l=6.dat};
			
			
			\addplot[smooth,thick,mark=square,color=pastelgreen,] table[ x expr=((\thisrowno{0})	), y expr=(\thisrowno{4})] {plots_data/scaledentanglement_l=3.dat};
			\addplot[smooth,thick,mark=square,color=pastelred,] table[ x expr=((\thisrowno{0})	), y expr=(\thisrowno{4})] {plots_data/scaledentanglement_l=4.dat};
			\addplot[smooth,thick,mark=square,color=pastelblue,] table[ x expr=((\thisrowno{0})	), y expr=(\thisrowno{4})] {plots_data/scaledentanglement_l=5.dat};
			\addplot[smooth,thick,mark=square,color=newcol,] table[ x expr=((\thisrowno{0})	), y expr=(\thisrowno{4})] {plots_data/scaledentanglement_l=6.dat};
			\addplot[domain=0:1,very thick,dashed,color=black]{1};		
		\end{axis}
	\end{tikzpicture}}
\includegraphics{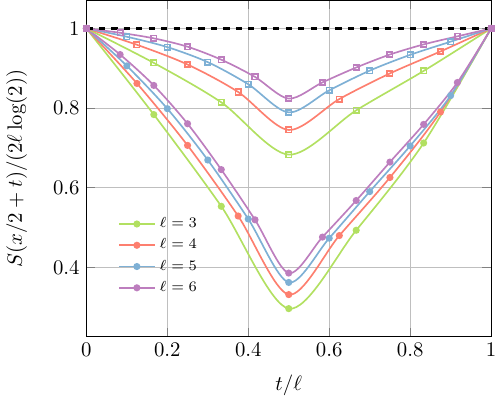}
	\caption{Numerical simulation of the evolution of the entanglement entropy in the time interval not fixed by dual unitarity, i.e.\ $t\in [x/2,\ell+x/2]$, for chaotic, dual unitary gates acting on qubits ($d=2$). The plot considers different values $\ell$ keeping the ratio ${x}/{\ell}$ constant (thus providing a scaling limit). The circle marks correspond to the choice ${x}/{\ell}=1$, while the squares correspond to the choice ${x}/{\ell}=4$. In both cases for increasing $x$ and $\ell$ we approach the straight line (black dashed)  describing the $x\rightarrow\infty$ limit.}
	\label{fig:plot2}
\end{figure}
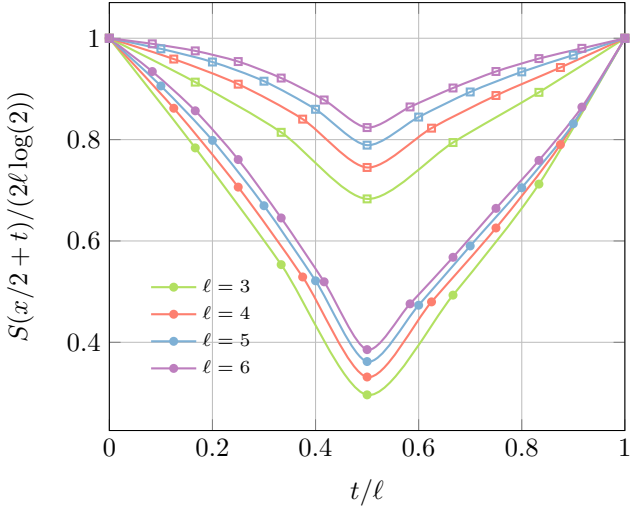

Theorem~\ref{thm:thm1} guarantees that for generic DU circuits with gates, say, in $\mathcal S_{\lt}$, the diagram in Fig.~\ref{fig:rhoA2} features a matrix $ T_{\ell}^\lt$ fulfilling 
\be
\quad  \left(T_{\ell}^\lt\right)^x \mapsto d^x \left[\frac{\ketbra{\mcirc}}{\braket{\mcirc}}\right]^{\otimes \ell}\!\!\!\!\!\!+O(\lambda_2^x), \quad 
\lambda_2 < d, \label{eq:replacementchaos}
\ee
for large enough $x$ ($O(\lambda_2^x)$ denotes an operator with norm scaling as $\lambda_2^x$). Making this replacement allows one to fully contract the diagram using Eq.~\ref{eq:DUgraph}  to obtain
\be
\rho_A(t)\propto \mathds{1}_{4\ell} + O(({\lambda_2}/{d})^x),\quad \lambda_2/d<1.
\ee
This means that also in the regime $\smash{x/2<t<x/2+\ell}$ the reduced state is proportional to the infinite temperature state. We then have
\be
\label{eq:membraneresult}
\!\!\!S_A(t)=\min(8t,4{\ell})\log(d) \!+\! O((\lambda_2/d)^x),\,\,\, \lambda_2/d<1,
\ee
for all times and interval lengths. For $x\gg 1$ this agrees with the membrane-picture prediction~\cite{nahum2017quantum}. We emphasise that to prove this statement we did not use any property of $T_{\ell}^\rt$, we only used that $T_{\ell}^\lt$ fulfils Eq.~\eqref{eq:replacementchaos}. We also note that the theorem does not guarantee that Eq.~\eqref{eq:membraneresult} also holds in the scaling limit, where $x\gg 1$ but $x/\ell$ is fixed. Indeed, in this limit the leading behaviour of the diagram in Fig.~\ref{fig:rhoA2} is not only specified by the leading eigenvalues of $T_{\ell}$. Our numerical experiments, however, suggest that Eq.~\eqref{eq:membraneresult} continues to hold: see Fig.~\ref{fig:plot2} for a representative example. 

Another immediate application of Theorem~\ref{thm:thm1} is to dynamical correlations of traceless operators with finite support in the infinite temperature state~\cite{bertini2019exact}. Recalling that $T_{z}^{\rt/\lt}$ are precisely the quantum channels characterising these correlations~\cite{bertini2019exact, claeys2021ergodic}, Theorem~\ref{thm:thm1} implies that for generic gates in $\mathcal S_{\lt}$ ($\mathcal S_{\rt}$) left (right) moving correlations of any operator with finite support decay to 0 at large times. In particular, since $\mathcal S_{\lt}=\mathcal S_{\rt}$ for $d=2$, this means that in \emph{almost all} dual-unitary circuits with $d=2$ all correlations decay to 0 at large times. Namely these systems are \emph{almost certainly} ergodic and mixing \emph{for all operators with finite support}.

Let us now prove Theorem~\ref{thm:thm1}.  We begin by writing 
\be
T_z^\lt =T_{z,0}^\lt\left(v \otimes v^*\right)^{\otimes z},
\label{eq:decomposition}
\ee
where $v\in U(d)$ (cf.\ Eq.~\eqref{eq:DUl}) and $T^{\lt}_{z,0}$ is the matrix generated by a gate in $\mathcal S_{\lt}$ with $v=\mathds{1}_d$. The latter is diagonal in the computational basis (its open legs correspond to the control inputs of the local gate) and thus we can directly characterise its spectrum. Specifically, the eigenvectors of  $T^{\lt}_{z,0}$ are written as 
\begin{align}
	\ket{\vec{j},\vec{k}}=\bigotimes_{a=1}^z \ket{j_a}\otimes\ket{k_a}, \qquad j_a,k_a=1,\ldots ,d \,,
	\label{eq:eigvec}
\end{align}
where $\ket{j}$ is the $j$th vector in the computational basis defining the gate, as  in Eq. \eqref{eq:controlright}.
The corresponding eigenvalues read as
\be
\lambda_{\vec{j},\vec{k}} =e^{{\rm{i}}\sum_{a}\phi_{j_a}-\phi_{k_a}}\tra{\rho^{\vec{j},\vec{k}}},
\ee
with 
\be
\rho^{\vec{j},\vec{k}} \equiv\rho^{[j_1]}\ldots \rho^{[j_z]}\smash{(\rho^{[k_1]}\ldots \rho^{[k_z]})}^\dagger,
\ee
and we set $\smash{u^{(j)}= e^{{\rm{i}} \phi_{j}} \rho^{(j)}}$ with $\smash{\phi_{j}\in\mathbb R}$ and $\smash{\rho^{(j)}\in SU(d)}$. Since $\smash{\rho^{\vec{j},\vec{k}}\in SU(d)}$ we have $\abs{\lambda}\le d$. 

To saturate the bound, one should have $\rho^{\vec{j},\vec{k}}= \alpha \mathds{1}_d$, with $\alpha^d=1$ to have unit determinant. This means that we should require
\be
\left(\rho^{\vec{j},\vec{k}}\right)^d =\mathds{1}_d. 
\label{eq:eq1}
\ee
We now recall that any random $z-$tuple of matrices from $SU(d)$ generate almost surely (according to the Haar measure) a \emph{free group} ~\cite{Epstein1971AlmostAS}. The free group property (see, e.g., Ref.~\cite{magnus2004combinatorial} for a formal definition) means that the only products of matrices built using $\{\rho^{[i]}\}$ and $\{\rho^{[j]^\dagger}\}$ that are equal to the identity, have to be simplifiable using unitarity. Namely, for Eq.~\eqref{eq:eq1} to hold we must have $\vec{j}=\vec{k}$. Noting then that $T_{z,0}^\lt$ is also \emph{normal} (this can be seen noting that the eigenvectors in Eq.~\eqref{eq:eigvec} are orthonormal and therefore $T_{z,0}^\lt$ can be diagonalised by a unitary matrix) we can then decompose it in orthogonal blocks as  
\begin{align}
	T_{z,0}^\lt =\left( d P_{\text{\rm max}}+T_{\text{rem}}\right),
	\label{eq:T0final}
\end{align}
where 
\begin{align}
	P_{\rm {\rm max}}=\sum_{\vec{j}\in \mathds{Z}_d^z}\ketbra{\vec{j},\vec{j}}=\Bigl(\sum_{j=1}^d \ketbra{j,j}\Bigr)^{\otimes z}\!\!\!\!\equiv p_{\rm {\rm max}}^{\otimes z},
\end{align}
is the projector on the block with eigenvalue $d$, while $\norm{T_{\text{rem}}}_{2}=\sup_{\ket{v}}{\norm{T_{\text{rem}} \ket{v}}}/{\norm{\ket{v}}}<d$~\footnote{Since the matrix is normal, its singular values are the square norm of eigenvalues}. 

Using the decomposition \eqref{eq:T0final} of $T_{z,0}^\lt $ in Eq.~\eqref{eq:decomposition} and defining the matrix  obtainined by applying the unitaries $v$ on $p_{\max}$ 
\begin{align}
		w = p_{\rm {\rm max}} (v \otimes v^*) p_{\rm {\rm max}}, 
\end{align}
one can readily show (see App.~\ref{sec:prooflemma1})     
\begin{lemma}
\label{lemma:lemma1}
The spectral radius of $T_{z}^\lt$ is bounded by $d$ and if there exists $\bra{\Psi}$, s.t., $\bra{\Psi} T_{z}^\lt = d e^{i \theta} \bra{\Psi}$ then
\be
\bra{\Psi}   w^{\otimes z}  \equiv   \bra{\Psi} \left(p_{\rm {\rm max}} (v \otimes v^*) p_{\rm {\rm max}} \right)^{\otimes z}  = e^{i \theta} \bra{\Psi} \,. 
\label{eq:wmatrix}
\ee
\end{lemma}
This means that any maximal left eigenvector of $T^\lt_z$,  $\bra{\Psi}$, must also be a left eigenvector of $w^{\otimes z}$ corresponding to an eigenvalue with magnitude one (which is maximal as it corresponds to the operator norm of $w$). We now note that in the orthonormal eigenbasis of $p_{\rm{\rm max}}$, the matrix $w$ has elements 
\begin{align}
	w_{ij}= \mel{i,i}{v\otimes v^*}{j,j}=\abs{v_{ij}}^2,
\end{align}
but, because $v$ is generic, almost surely it has 
\begin{align}
	v_{ij}\ne 0 \qquad \forall i,j=1,\ldots d\implies \abs{v_{ij}}^2>0.
\end{align}
Therefore, the matrix $w$ has almost certainly strictly positive entries and the Perron--Frobenius Theorem~\cite{gantmacher1980theory} guarantees that it has a unique eigenvalue with strictly maximal magnitude. Since one can immediately verify $\bra{\mcirc} w=\bra{\mcirc}$, we then must have  $\smash{\bra{\Psi}\propto\bra{\mcirc}^{\otimes z}}$.\qedsymbol{}

\subsection{Charged Dual-Unitary Circuits}
\label{sec:entanglementcharged}

Let us now consider dual-unitary circuits with commuting $U(1)$ charges. As shown in Ref.~\cite{foligno2024nonequilibrium} in these circuits one can arrange the charge densities to be mutually orthogonal projectors supported on one site, $\{\Pi^{\lt/\rt}_{\alpha}\}_\alpha$, and fulfilling 
\begin{align}
		\sum_{\alpha}\Pi^{\lt/\rt}_{\alpha}=\mathds{1}_d, \qquad 
	U(\Pi^{\rt}_\alpha\otimes\Pi^{\lt}_\beta )=(\Pi^{\lt}_\beta \otimes\Pi^{\rt}_\alpha) U.
	\label{eq:solitoncondition}
\end{align}
The second equation means that $\{\Pi^{\lt/\rt}_{\alpha}\}_\alpha$ behave as ``solitons''~\cite{bertini2020operator2}: time evolution just shifts them to the left ($\lt$) or to the right ($\rt$) but does not modify them. The local gate can then be \emph{decomposed} into smaller dual unitary blocks, acting on qudits of dimensions $\smash{d^{\rt/\lt}_\alpha\equiv \tra{\Pi^{\lt/\rt}_{\alpha}}}$
\be
 	U^{\alpha,\beta}= \left(\Pi^\lt_\beta \otimes \Pi^\rt_\alpha\right) U \left(\Pi^\rt_\alpha  \otimes \Pi^\lt_\beta \right),
\label{eq:blockdef} 
\ee
and the expectation value of $\{\Pi^{\lt/\rt}_{\alpha}\}_\alpha$ on the initial state defines a classical probability of being in the sectors $\alpha,\beta$, i.e., 
\be
\smash{c_{\alpha,\beta}=\tra{\Pi^{\lt}_\alpha\Pi^{\rt}_\beta}/d\ge 0}.
\ee 


Our starting point is again the density matrix $\rho_A$ in Fig.~\ref{fig:rhoA2}. The idea is to define suitable quantum channels in terms of  $\{\Pi^{\lt/\rt}_{\alpha}\}_\alpha$ and apply them to $\rho_A$. Then we use the monotonicity of the quantum relative entropy \cite{nielsen2010quantum} to bound $S[\rho_A]$. As explicitly shown in Appendix~\ref{sec:app1} this gives   
\begin{theorem}
\label{thm:thm2}
For $\smash{x/2<t<x/2+\ell}$, the entanglement entropy fulfils 
\begin{align}
\label{eq:boundmutual}
	S[\rho_A] \leq 4{\ell} \log(d)-\left(\ell-\abs{2t-(x+\ell )}\right) I_{\lt:\rt},
\end{align}
where $I_{\lt:\rt}$ is the classical mutual information between left and right-moving charges according to the probability distribution $c_{\alpha,\beta}$. 
\end{theorem}
This shows that, whenever the number of conserved charges is larger than zero (and hence $I_{\lt:\rt}>0$), the entropy has a drop when the two intervals become causally connected (\emph{without} any large distance assumption, only requiring $x>\ell$), in agreement with the quasi-particle-picture prediction. In fact, reasoning as in the proof of Theorem~\ref{thm:thm1} we can establish (see Appendix~\ref{sec:app2})
\begin{theorem}
\label{thm:thm3}
If each block $U^{\alpha,\beta}$ has local dimension $d^{\lt/\rt}_{\alpha}=2$ and is generic (and chosen independently from the others) the bound in Eq.~\eqref{eq:boundmutual} \emph{is saturated} for $x\gg \ell$. 
\end{theorem}

\section{Conclusions}
\label{sec:conclusions}

In this work we showed that in dual unitary circuits the growth of entanglement of two disjoint intervals depends on the nature of the microscopic dynamics. Specifically, \emph{generic} dual-unitary circuits follow the entanglement membrane picture put forward in Ref.~\cite{jonay2018coarsegrained, zhou2020entanglement}, while \emph{charged} dual-unitary circuits, follow the quasiparticle picture of Ref.~\cite{calabrese2005evolution}. Interestingly, this is the case despite the fact that charged dual unitary circuits are generically not Yang--Baxter-integrable~\cite{foligno2024nonequilibrium} (see also Ref.~\cite{claeys2024operator}). Our results also led to the proof that generic dual-unitary circuits with $d=2$ have almost certainly ergodic and mixing dynamical correlation functions for \emph{all operators of finite support}.

To the best of our knowledge the one presented here is the first rigorous result confirming the validity of the entanglement membrane picture for clean, microscopic systems and in a setting where it differs qualitatively from that of the quasiparticle picture. On the other hand, we also found that charged dual-unitary circuits do not follow the membrane picture despite not having conserved quasiparticles. Instead, they follow the quasiparticle picture. This is not the first observation that conservation laws invalidate the membrane picture~\cite{rakovszky2019sub, huang2020dynamics}, and calls for for future research on a systematic characterisation of quantum many-body dynamics in non-integrable systems with conservation laws.

More generally, one of the main ingredients of our proofs has been the fact that $U(d)$ matrices generate almost surely a free group~\cite{Epstein1971AlmostAS}, which we used to characterise the spectra of the relevant transfer matrices. This general idea can be applied more widely to investigate other probes of the non-equilibrium dynamics --- like OTOCs and operator entanglement --- or spectral statistics.

\begin{acknowledgements}
We thank Paolo Tognini for very fruitful discussions and insights on the technical aspects of this work. This research has been supported by the Royal Society through the University Research Fellowship No. 201101. We warmly acknowledge the hospitality of the Simons Center for Geometry and Physics during the program “Fluctuations, Entanglements, and Chaos: Exact Results” and the International School for Advanced Studies (SISSA) where part of this work has been performed.
\end{acknowledgements}

\appendix


\section{Proof of Lemma~\ref{lemma:lemma1}}
\label{sec:prooflemma1}

To prove the first part of Lemma~\ref{lemma:lemma1} we note that for any normalised vector $\ket{\Psi}$
\begin{align}
\mel{\Psi}{ T^{\lt}_z \left(T^{\lt}_z\right)^\dagger }{\Psi}= \mel{\Psi}{ T^{\lt}_{z,0}  \left(T^{\lt}_{z,0}\right)^\dagger  }{\Psi} \leq d^2
\label{eq:norm}
\end{align}
where in the second step we used Eq.~\eqref{eq:decomposition} and the fact that the spectral radius of $T^{\lt}_{z,0}$ (which coincides with its operator norm since the matrix is normal) is $d$. Therefore both the operator norm of $T^{\lt}_z$ and its spectral radius (bounded by the operator norm) are bounded by $d$. 

To prove the second part, let $\bra{\Psi}$ be a maximal left eigenvector of $T^\lt$
\be
\bra{\Psi} T_{z,0}^\lt = d e^{i \theta}  \bra{\Psi} .
\label{eq:eigenvalueprop}
\ee
Then Eq.~\eqref{eq:norm} implies 
\be\mel{\Psi}{
T^{\lt}_{z,0} \left(T^{\lt}_{z,0}\right)^\dagger }{\Psi}  = d^2 . 
\ee
Recalling the decomposition \eqref{eq:T0final}, this immediately implies that $\bra{\Psi}$ must belong to the orthogonal block of $T^\lt_{z,0}$ corresponding to the largest eigenvectors, i.e. $P_{\text{\rm max}}$
\be
\bra{\Psi} P_{\text{\rm max}}=\bra{\Psi} \qquad \bra{\Psi} T_{\text{rem}}=0. \label{eq:neweq}
\ee 
Using Eqs.~\eqref{eq:decomposition}, \eqref{eq:T0final} and \eqref{eq:eigenvalueprop} we then find 
\be
\bra{\Psi}  \left(v \otimes v^*\right)^{\otimes z} =  e^{i \theta} \bra{\Psi}\,.
\ee
Finally, using multiple times the first relation of Eq. \eqref{eq:neweq}, we can rewrite the last equation as \begin{align}
	\bra{\Psi}P_{\text{\rm max}}\left(v \otimes v^*\right)^{\otimes z}P_{\text{\rm max}}= e^{i \theta} \bra{\Psi},
\end{align}
which is Eq.~\eqref{eq:wmatrix} of the main text.

\section{Proof of Theorem~\ref{thm:thm2}}
\label{sec:app1}

In this section, we show more in details how dual unitarity and charge conservation allow us to bound the entanglement entropy of the reduced density matrix reported in Fig. \ref{fig:rhoA} of the main text, proving Theorem \ref{thm:thm2}.
	
	The idea is to apply suitable unital quantum channels to the reduced density matrix. A unital quantum channel is a completely positive, trace preserving map, which leaves the identity matrix invariant. The monotonicity of the quantum relative entropy \cite{nielsen2010quantum} implies that a unital quantum channel can only increase the entanglement entropy. Specifically, we apply the following quantum channel
\begin{align}
		\mathcal{E}^\lt[X]\equiv\sum_{\alpha=1}^{m^\lt}\Pi^\lt_\alpha\frac{\tra{\Pi^\lt_\alpha X}}{d^\lt_\alpha},
		\label{eq:lchannel}
\end{align}
on the left-pointing legs on the left interval of the reduced density matrix (see Fig. \ref{fig:rhoA}) and the channel 
\begin{align}
		\mathcal{E}^\rt[X]\equiv\sum_{\alpha=1}^{m^\lt}\Pi^\rt_\alpha\frac{\tra{\Pi^\rt_\alpha X}}{d^\lt_\alpha},
\end{align}
on the right-pointing legs on its right interval.
	
First, we show that $\mathcal{E}^{\lt/\rt}[X]$ is indeed a quantum channel. We can choose an orthonormal basis for the local Hilbert space on each leg, $\ket{i,\alpha}_{i=1,\ldots, d^\lt_\alpha}$, which is compatible with the orthogonal projectors $\Pi^\lt_\alpha$, i.e., 
\begin{align}
\sum_{i=1}^{d^\lt_\alpha}\ketbra{i,\alpha}=\Pi^\lt_\alpha.
\end{align}
Then, we can write Eq.~\eqref{eq:lchannel} in Kraus form
\begin{align}
\mathcal{E}^\lt[X]=\sum_{\alpha=1}^{m^\lt}\sum_{i,j=1}^{d^\lt_\alpha}\ketbra{i,\alpha}{j,\alpha}{X}\ketbra{j,\alpha}{i,\alpha}\frac{1}{{d^\lt_\alpha}}, \label{eq:lchannelkraus}
\end{align}
proving that $\mathcal{E}^\lt[X]$ is indeed a completely positive map. The channel is also trace preserving, as can be seen from the following chain of equalities
\begin{align}
		\tra{\mathcal{E}^\lt[X]}&=\sum_{\alpha=1}^{m^\lt}{\tra{\Pi^\rt_\alpha X}}\notag\\
		&={\tra{\left(\sum_{\alpha=1}^{m^\lt}\Pi^\rt_\alpha \right)X}}=\tra{X},
\end{align}
where we used the first statement in Eq. \eqref{eq:solitoncondition} for the last equality. Finally, $\mathcal{E}^\lt[X]$ is \emph{unital}, meaning it preserves the identity
\begin{align}
		\mathcal{E}^{\lt}[\mathds{1}_d]=\sum_{\alpha=1}^{m_\lt} \Pi^\lt_\alpha \frac{\tra{\Pi^\lt_\alpha}}{d^\lt_\alpha}=\sum_{\alpha=1}^{m_\lt} \Pi^\lt_\alpha =\mathds{1}_d,
\end{align}
where we used again Eq. \eqref{eq:solitoncondition}. We now introduce a graphical notation for the solitons $\Pi^{\rt/\lt}_\alpha$ in order to make the calculations more transparent
	\begin{align}
		\Pi_\alpha^{{\rt}} \otimes \mathds{1}_d = \fineq[-0.8ex][1.4][1]{	
			\draw[thick] (0,-.5)--++(0.5,0);
			\charge[0.25][-.5][red]}, \quad\quad \Pi_\alpha^{{\lt}} \otimes \mathds{1}_d =  \fineq[-0.8ex][1.4][1]{	
			\draw[thick] (0,.5)--++(.5,0);
			\charge[0.25][.5][blue]}.\label{eq:chargepic}
	\end{align}
	The soliton property \eqref{eq:solitoncondition} can be written as \begin{align}
		\fineq[-0.8ex][1][1]{
			\roundgate[0][0][1][topright][bertiniorange][-1]
			\charge[-.35][-.35][red]
		}=\fineq[-0.8ex][1][1]{
			\roundgate[0][0][1][topright][bertiniorange][-1]
			\charge[.35][.35][red]
		}, \qquad \quad\fineq[-0.8ex][1][1]{
			\roundgate[0][0][1][topright][bertiniorange][-1]
			\charge[.35][-.35][blue]
		}=\fineq[-0.8ex][1][1]{
			\roundgate[0][0][1][topright][bertiniorange][-1]
			\charge[-.35][.35][blue]
		}.
		\label{eq:Picharges}
	\end{align}
If we represent an operator $X$ in the replica space as \begin{align}
	\fineq[-0.8ex][1][1]{
		\draw(0,0)--++(0,1);
		\draw[thick,fill=white] (0,0) circle (0.3);	
		\node at (0,0) {$X$};
	},
\end{align}
then the action of the quantum channel \eqref{eq:lchannel} acting on it, can be represented as\begin{align}
	\mathcal{E}^{\lt}[X]=\sum_\alpha \fineq[-0.8ex][1][1]{
		\draw(0,0)--++(0,1);
		\draw[thick,fill=white] (0,0) circle (0.3);	
		\draw[thick] (0,1.5)--++(0,.5);
		\node at (0,0) {$X$};
		\cstate[0][1]
		\charge[0][0.6][blue]
		\charge[0][1.8][blue]
				\cstate[0][1.5]
		\draw[->](0.1,1.8)--(1,1.5);
		\draw[->](0.1,0.6)--(1,1.4);
		\node at (1.25,1.5) {$\Pi^\lt_\alpha$};
}\frac{1}{d^\lt_\alpha}.
\end{align}
	In this notation, the action of the quantum channels $\mathcal{E}^{\lt/\rt}$ on the reduced density matrix in Fig \eqref{fig:rhoA2} of the main text is represented as
	\begin{widetext}
	\begin{align}
\left(\mathcal{E}^\rt\right)^{\otimes {\ell}}\left(\mathcal{E}^\lt\right)^{\otimes {\ell}}[\rho_A]\propto		\sum_{\vec{\alpha},\vec{\beta}}	\frac{1}{d^\lt_{\vec{\alpha}}d^\rt_{\vec{\beta}}}\fineq[-0.8ex][0.65][1]{
			\foreach \i in {0,...,5}
			{
				\foreach \j in {0,...,1}
				{
					\roundgate[-\i+\j][\i+\j][1][topright][bertiniorange][-1]
					\roundgate[4+\i-\j][\i+\j][1][topright][bertiniorange][-1]
				}
			}
			\roundgate[-6][6][1][topright][bertiniorange][-1]
			\roundgate[10][6][1][topright][bertiniorange][-1]
			\roundgate[1][-1][1][topright][bertiniorange][-1]
			\roundgate[3][-1][1][topright][bertiniorange]-1]
			\roundgate[2][0][1][topright][bertiniorange][-1]
			\foreach \i in {-1,...,6}
			{
				\cstate[-\i-.5][\i-.5]
				\cstate[4+\i+.5][\i-.5]
			}
			\foreach \i in {0,...,4}
			{
				\cstate[-\i+1.5][\i+1.5]
				\cstate[4+\i-1.5][\i+1.5]
			}	
			\bellpair[2][-2]
			\draw[thick] (-6.5+15.25,6.5+.25)--++(.5,.5);	
			\draw[thick] (-6.5+17.25,6.5+.25)--++(.5,.5);	
			\draw[thick] (-8+1.25,6.5+.25)--++(-.5,.5);	
			\draw[thick] (-8+3.25,6.5+.25)--++(-.5,.5);	
			\charge[-6.3+2][6.3][blue]
			\charge[-6.3][6.3][blue]
			\charge[-6.3+2-.7][6.3+.7][blue]
			\charge[-6.3-.7][6.3+.7][blue]
			\charge[-6.2+2+14.5][6.3][red]
			\charge[-6.2+14.5][6.3][red]
			\charge[-6.2+2+14.5+.7][6.3+.7][red]
			\charge[-6.2+14.5+.7][6.3+.7][red]
			\cstate[-6.5+2][6.5]
			\cstate[-6.5][6.5]
			\cstate[-6.5+2][6.5]
			\cstate[-6.5-.25][6.5+.25]
			\cstate[-6.5+2-.25][6.5+.25]
			\cstate[-6.5+2+15][6.5]
			\cstate[-6.5+15][6.5]	
			\cstate[-6.5+2.25+15][6.5+.25]
			\cstate[-6.5+15.25][6.5+.25]
			\draw[blue,dashed,rounded corners=10pt] (-6,7)--++(8,-8)--++(8,8)--++(1,-1)--++(-9,-9)--++(-9,9)--cycle;
		}\!\!\!\!\!\!\!\!
		\label{eq:diagrampairs}
		\end{align}
	where, for simplicity, we dropped an overall normalisation factor (we can restore it at the end by imposing that the trace of the resulting matrix is $1$) and we summed over strings $\vec{\alpha},\vec{\beta}$ of $\ell$ projectors: $\vec{\alpha}=(\alpha_1,\ldots\alpha_\ell)$, and similarly for $\vec{\beta}$.
	The factors $d^\lt_{\vec{\alpha}},d^\rt_{\vec{\beta}}$ are a shorthand notation for the product of all the factors coming from the channels
\begin{align}
		d^\lt_{\vec{\alpha}}=\prod_{i=1}^\ell d^\lt_{\alpha_i}\qquad d^\rt_{\vec{\beta}}=\prod_{i=1}^\ell d^\rt_{\beta_i}.
\end{align}
	
	 Using the relation \eqref{eq:Picharges}, combined with dual unitarity \eqref{eq:unitarityfoldeddiagram}--\eqref{eq:DUgraph}, we can simplify the diagram \eqref{eq:diagrampairs} to obtain (ignoring global normalisation factors)
	\begin{align}
		 \sum_{\vec{\alpha},\vec{\beta}}\frac{1}{d^\lt_{\vec{\alpha}}d^\rt_{\vec{\beta}}}	\fineq[-0.8ex][0.65][1]{
			\draw[orange,dashed,rounded corners=10pt] (-8,7)--++(2,-2)--++(1,1)--++(-2,2)--cycle;
			\draw[blue,dashed,rounded corners=10pt] (-6,7.5)--++(5,0)--++(-2.5,-2.5)--cycle;
			\draw[orange,dashed,rounded corners=10pt] (1,7)--++(-2,-2)--++(-1,1)--++(2,2)--cycle;
			\draw[thick] (-6.5+4.25,6.5+.25)--++(.5,.5);	
			\draw[thick] (-6.5+6.25,6.5+.25)--++(.5,.5);	
			\draw[thick] (-8+1.25,6.5+.25)--++(-.5,.5);	
			\draw[thick] (-8+3.25,6.5+.25)--++(-.5,.5);	
			\draw[thick] (-6.5+4,6.5)--++(-.5,-.5);	
			\draw[thick] (-6.5+6,6.5)--++(-.5,-.5);	
			\draw[thick] (-8+1.5,6.5)--++(.5,-.5);	
			\draw[thick] (-8+3.5,6.5)--++(.5,-.5);	
			\charge[-6.3+2][6.3][blue]
			\charge[-6.25][6.25][blue]
			\charge[-6.3+2-.7][6.3+.7][blue]
			\charge[-6.3-.7][6.3+.7][blue]
			\charge[-6.2+2+3.45][6.25][red]
			\charge[-6.2+3.5][6.3][red]
			\charge[-6.2+2+3.5+.7][6.3+.7][red]
			\charge[-6.2+3.5+.7][6.3+.7][red]
			\cstate[-6.5+2][6.5]
			\cstate[-6.5][6.5]
			\cstate[-6][6]
			\cstate[-1][6]
			\cstate[-6.5+2][6.5]
			\cstate[-6.5-.25][6.5+.25]
			\cstate[-6.5+2-.25][6.5+.25]
			\cstate[-6.5+2+4][6.5]
			\cstate[-6.5+4][6.5]	
			\cstate[-6.5+2.25+4][6.5+.25]
			\cstate[-6.5+4.25][6.5+.25]
			\bellpair[-3.5][5.5]
			\draw[thick] (-6.5-2.25,6.75)--++(.5,.5);
			\draw[thick] (-6.5-4.25,6.75)--++(.5,.5);
			\draw[thick] (-6.5+8.25,6.75)--++(-.5,.5);
			\draw[thick] (-6.5+10.25,6.75)--++(-.5,.5);
			\cstate[-6.5+4.25][6.5+.25]
			\cstate[-6.5-2.25][6.75]
			\cstate[-6.5-4.25][6.75]
			\cstate[-6.5+8.25][6.75]
			\cstate[-6.5+10.25][6.75]
		}\label{eqboh}
	\end{align}
	The number of pairs of charges connected via an initial state is $\ell-\abs{2t-(x+\ell)}$ (as the one circled in blue in \eqref{eq:diagrampairs}--\eqref{eqboh}). The remaining charges, which are not connected, (see e.g. the ones circled in orange in Eq. \eqref{eqboh}) can be resummed to the infinite temperature state. Focusing on a circled orange leg, notice that the factor ${1}/{d^{\lt/\rt}_{\vec{\alpha}}}$ gets cancelled by the matrix element of the projectors between two bullet states, which is just
\begin{align}
\fineq{
		\draw[thick] (0,0)--++(-.5,.5);		
		\cstate[0][0]
		\cstate[-.5][.5]
		\charge[-.25][.25][blue]
}=	\tra{\Pi^{	\lt}_\alpha}=d^{\lt}_\alpha \qquad \fineq{
\draw[thick] (0,0)--++(.5,.5);		
\cstate[0][0]
\cstate[.5][.5]
\charge[.25][.25][red]
}=	\tra{\Pi^{	\rt}_\alpha}=d^{\rt}_\alpha.
\end{align}	
	  Then, using the fact that\begin{align}
		\sum_{\alpha} \fineq{
			\draw[thick] (0,0)--++(-.5,.5);		
			\cstate[0][0]
			\charge[-.25][.25][blue]
		} = \sum_{\alpha} \Pi^{\lt}_{\alpha}=\mathds{1}_d=\fineq{
		\draw[thick] (0,0)--++(-.5,.5);		
		\cstate[0][0]
	} ,
	\end{align}
it is clear that the system is in the infinite temperature state on such legs. Putting everything together, after the channel, the (normalised) density matrix can be written as 
\begin{align}
		\rho=\left(\frac{\mathds{1}_d}{d}\right)^{\otimes 2\ell+\abs{4t-2(x+\ell)}} \otimes \left(\sum_{\alpha,\beta} \Pi^\lt_\alpha\otimes \Pi^\rt_\beta \frac{\tra{\Pi^\lt_\alpha\Pi^\rt_\beta}}{d d^\lt_\alpha d^\rt_\beta}\right)^{\otimes \ell-\abs{2t-(x+\ell)}},
\end{align}
\end{widetext}
whose entanglement entropy can be written in terms of the classical probability defined  a single pair of the initial state
\begin{align}
c_{\alpha,\beta}\equiv\frac{\tra{\Pi_\alpha^\lt\Pi_\beta^\rt}}{d}.
\end{align}
The entanglement entropy after the channel is then
\begin{widetext}
\begin{align}
S[\left(\mathcal{E}^\lt\right)^{\otimes \ell}\left(\mathcal{E}^\rt\right)^{\otimes \ell}[\rho_A]]=		\left(2\ell+\abs{4t-2(x+\ell)}\right) \log(d)+\left(\ell-\abs{2t-(x+\ell)}\right)\sum_{\alpha,\beta} c_{\alpha,\beta}\left(\log(d_\alpha d_\beta)-\log(c_{\alpha,\beta})\right).\label{eq:exprS1}
\end{align}
This expression can be rewritten by defining the marginals of $c_{\alpha,\beta}$, which obey
\begin{align}
c^\lt_\alpha=\sum_{\beta} c_{\alpha,\beta}=\frac{\tra{\Pi^\lt_\alpha}}{d}=\frac{d^\lt_\alpha}{d} \qquad c^\rt_\beta=\sum_\alpha c_{\alpha,\beta}=\frac{d^\rt_\beta}{d},
\label{eq:entdrop}
\end{align}
After simple manipulations we obtain 
\begin{align}
	S[\left(\mathcal{E}^\lt\right)^{\otimes \ell}\left(\mathcal{E}^\rt\right)^{\otimes \ell}[\rho_A]]=4\ell \log(d)-(\ell-\abs{2t-(x+\ell)}\left[\sum_{\alpha,\beta} c_{\alpha,\beta} \log(c_{\alpha,\beta})-\sum_{\alpha} c^\lt_\alpha\log(c^\lt_\alpha)-\sum_{\beta} c^\rt_\beta\log(c^\rt_\beta)\right].\label{eq:entanglementexpression}
\end{align}
\end{widetext}
The quantity in square brackets in \eqref{eq:entanglementexpression} corresponds to the classical mutual information of the probability distribution $c_{\alpha,\beta}$, implying that it is strictly positive. Expression \eqref{eq:entanglementexpression}, which only depends on the expectation value of the solitons $c^{\lt/\rt}_\alpha$ and $c_{\alpha,\beta}$, generalises to any charged solvable state, as defined in \cite{foligno2024nonequilibrium}: here we considered the simplest case to make the calculations more transparent.

 \section{Proof of Theorem~\ref{thm:thm3}}
 \label{sec:app2}
 
 To prove Theorem~\ref{thm:thm3} we consider a transfer matrix built with $m^2$ blocked dual unitary gates  acting on qudits of local dimension $d=2m$ 
\begin{align}
	U=\bigoplus_{\alpha=1}^{m}\bigoplus_{\beta=1}^{m} U^{\alpha,\beta};
\end{align}
\begin{align}
	U^{\alpha,\beta}=U^{\alpha,\beta}(\Pi^\rt_{\alpha}\otimes \Pi^\lt_\beta)=(\Pi^\lt_\beta\otimes \Pi^\rt_{\alpha})U^{\alpha,\beta}
\end{align}
where the blocks act on qubits (i.e. on a local dimension $d_{loc}={d}/{m}=2$)\begin{align}
	d^{\lt/\rt}_\alpha=\tra{\Pi^{\lt/\rt}_\alpha}=2\qquad d=2m^\rt=2m^\lt.
\end{align} 
The blocks are assumed to be generic dual unitaries chosen independently from $\mathcal S_{\lt}$ or $\mathcal S_{\rt}$ for each $\alpha,\beta$. Since both the parametrization in Eq \eqref{eq:DUl} and the one obtained by spatial reflection  are complete and coincide in $d=2$, we can use either depending on our convenience. We start by using a parametrization \eqref{eq:DUl} and consider left-transfer matrices.\\
The gate in a charge block $U^{\alpha,\beta}$ can be written as \begin{align}
	U^{\alpha,\beta}=S \cdot U^{[u],\alpha,\beta,\lt} \cdot\left(v^{\alpha,\beta} \otimes \mathds{1}_2\right) \qquad v^{\alpha,\beta}\in SU(d)
\end{align}
where the matrices $u^{\alpha,\beta}\equiv\rho^{(i)}_{\alpha,\beta}e^{\phi_{i,\alpha,\beta}}$, which define a control gate as in \eqref{eq:controlright}, are drawn randomly and independently for each $i,\alpha,\beta$.
Given the block structure, the left transfer matrix can be written as a sum of other transfer matrices $T^\lt_{z,\alpha}$ which have their main diagonal projected on the right-moving soliton $\Pi^\rt_\alpha$
\begin{align}
	T^\lt_z=\sum_{\alpha=1}^{m_{\rt}} T^\lt_{z,\alpha},
\end{align}
the transfer matrices $T^\lt_\alpha$ can be represented using the notation introduced the previous section for the solitons
\begin{align}
	T^\lt_{z,\alpha}=\fineq[-0.8ex][.75]{
		\def \size {3}
		\foreach \i in {1,...,\size}
		{
			\roundgate[\i][\i][1][topleft][bertiniorange][-1]
			\cstate[.5][.5]
			\cstate[\size+.5][\size+.5]
		}
		\draw[thick,decorate,decoration={brace}] (.15,1.25)--++(\size-0.5,\size-0.5);
		\charge[3.3][3.3][red]
		\node[scale=1.5] at (3.1,3.7) {$\alpha$};
		\node[scale=1.5] at (1.,2.75) {$z$};
	}.
\end{align}
Due to the unitarity of the folded gate, it is immediate to see that each transfer matrix $T^\lt_\alpha$ has eigenvalues of modulus $\le d^{\rt}_\alpha=2$.
We now show that \emph{each} $T^\lt_{z,\alpha}$ has the same set of largest eigenvectors, all with eigenvalue $2$, thus implying that these are the largest eigenvectors of the sum $T^\lt_z=\sum_\alpha T^\lt_{z,\alpha}$.
\\
Thanks to the soliton conservation condition \eqref{eq:solitoncondition}, the transfer matrices commute with all the strings of left-moving projectors
\be
\Pi^\lt_{\vec{\beta},\vec{\gamma}}=\bigotimes_{i=1}^z \Pi^{\lt}_{\beta_i}\otimes \Pi^{\lt}_{\gamma_i},
\ee
namely
\be
[T^{\lt}_{z,\alpha}, \Pi^\lt_{\vec{\beta},\vec{\gamma}}]=0,
\ee
where $\Pi^\lt_{\beta_i}$ is understood to act on the first layer of the space on which $T^\lt_z$ is defined (the ``forward'' copy), whereas $\Pi^\lt_{\gamma_i}$ acts on the complex conjugate layer (or the ``backward'' copy). The index $i$ just specifies the spatial position of the leg of the transfer matrix considered. Thanks to this property, we can look at the spectrum of the transfer matrix projected on  a reduced space\begin{align}
\Pi^\lt_{\vec{\beta},\vec{\gamma}}	T^{\lt}_{z,\alpha}\Pi^\lt_{\vec{\beta},\vec{\gamma}}.\label{eq:blockT}
\end{align}
In each block, the transfer matrix is now built with generic $DU$ gates of local dimension $d=2$. By using the reasoning of the main text, first setting $v^{\alpha,\beta}=\mathds{1}_{2}$ shows that the eigenvalues of a transfer matrix can be written as strings of bits $\ket{\left(\vec{j},\vec{\beta}\right),\left(\vec{k},\vec{\gamma}\right)}$, where\begin{align}
\ket{\left(\vec{j},\vec{\beta}\right),\left(\vec{k},\vec{\gamma}\right)}\equiv \prod_{a=1}^z	\ket{\left(j_a,\beta_a\right)}\otimes\ket{\left(k_a,\gamma_a\right)},
\end{align}
with $j_a, k_a=1,2$, where it is understood that$\{\ket{(j,\beta)}\}$  is an orthonormal basis of the Hilbert space projected by $\Pi^\lt_\beta$.
The corresponding eigenvalue of $\ket{\left(\vec{j},\vec{\beta}\right),\left(\vec{k},\vec{\gamma}\right)}$ is
\begin{widetext}
\begin{align}
	\lambda_{\vec{j},\vec{\beta},\vec{k},\vec{\gamma}}=
	\exp(\sum_{a=1}^z \phi_{{j_a},\alpha,\beta_a}-\phi_{{k_a},\alpha,\gamma_a})\tra{\left(\prod_{a=1}^z \rho^{(j_a)}_{\beta_a,\alpha}\right)\left(\prod_{a=1}^z \rho^{(k_a)}_{\gamma_a,\alpha}\right)^\dagger}. 
\end{align}
\end{widetext}
In order to have a maximal eigenvalue, the matrix inside the trace must be the identity, but because of the free group property (cf.\ Ref.~\cite{Epstein1971AlmostAS}) this can happen only if 
\begin{align}
	j_a=k_a,\qquad \beta_a=\gamma_a. 
\end{align}
Therefore, the only charge blocks of the transfer matrix in Eq.~\eqref{eq:blockT} with maximal eigenvalue must have the same charges on both the forward and backward layer of the gates (i.e. $\vec{\beta}=\vec{\gamma}$). Once we restrict to one such reduced space, we choose the matrices $v_{\alpha,\beta}$, independently in each block. We can then use Perron-Frobenius Theorem to show that in each of these block there can only be a simple largest eigenvalue, which is the bullet state (defined with the states in that block). 
This shows that the eigenvectors can be written as  
\begin{align}
\ket{\psi^{\rm max}_\vec{\beta}}=	\bigotimes_a \sum_{j=1}^2 \ket{(j,\beta_a),(j,\beta_a)}
\end{align}
with $j_a=1,2 , \beta_a=1, \ldots m^\lt$. Large powers of the transfer matrix will project on its leading eigenspace, thus we can write 
\begin{widetext}
\begin{align}
	\lim_{x\rightarrow\infty} \left(T^{\lt}_z\right)^x=\sum_\vec{\beta} \bigotimes_{a=1}^z \sum_{i,j=1}^2 \ketbra{(j,\beta_a)(j,\beta_a)}{(i,\beta_a)(i,\beta_a)}=\left[\sum_{\beta=1}^m \sum_{i,j=1}^2 \ketbra{(j,\beta)(j,\beta)}{(i,\beta)(i,\beta)}\right]^{\otimes z}.
	\label{eq:largepower}
\end{align}
\end{widetext}
The transfer matrix $T^\lt_x$ is a quantum channel acting on operators defined on the original Hilbert space with local dimension $d=2m$; the last term in Eq. \eqref{eq:largepower} can be recognized to be the folded version of the channel $\left(\mathcal{E}^\lt\right)^{\otimes z}$ defined in the previous section (cfr. Eq. \eqref{eq:lchannelkraus}). Therefore we have
\begin{align}
	\lim_{x\rightarrow\infty} \left(T^{\lt}_z\right)^x=\mathcal{E}^\lt;
\end{align}
the same reasoning applies also to right transfer matrices
\begin{align}
	\lim_{x\rightarrow\infty} \left(T^{\rt}_z\right)^x=\mathcal{E}^\rt,
\end{align}
showing saturation of the bound obtained in Sec. \ref{sec:app1} for large values of the distance between the two intervals $x$.

\bibliography{bibliography.bib}
\bibliographystyle{quantum}

\end{document}

%% file: tikz_lib.tex
\usepackage{standalone, gensymb, tensor, amsmath,amsfonts,amssymb,amsthm,bbm,bm, braket}
\usepackage[T1]{fontenc}
\usepackage{hyperref}
\usepackage[inline]{enumitem}
\usepackage{scalerel, physics}
\usepackage{wasysym}
\usepackage{comment}
\usepackage{enumerate}
\usepackage{graphicx}
\usepackage{mathtools}
\usepackage{soul}
\usepackage{xparse}
\usepackage{dsfont}
\usepackage{ifthen}
\usepackage{tensor}
\usepackage{tikz}
\usepackage{pgfplots}
\usepackage{tikz-network}
\usepackage{simplewick} 
\usepackage{ mathrsfs }

\usetikzlibrary{patterns,decorations.pathreplacing}
\theoremstyle{break}        

\theoremstyle{break}

\usepackage{color}
\definecolor{alessandrored}{RGB}{250, 153, 140}
\definecolor{OliveGreen}{RGB}{85,107,47}
\definecolor{NavyBlue}{RGB}{0,0,128}
\definecolor{alessandrogreen}{RGB}{17, 127, 35}
\definecolor{jonasgreen}{RGB}{81, 160, 37}
\usepackage{tensor}
\usepackage{xifthen}
\usepackage{xargs}

\newcommandx{\fineq}[5][1=-.8ex,2=1,3=1,5=0]{
	\begin{tikzpicture}[baseline={([yshift=#1]current  bounding  box.center)}, scale = #2, every node/.style={scale = #3},rotate around={#5:(0,0)},every node/.style={transform shape}]
		#4
	\end{tikzpicture}
}

\definecolor{bertinired}{RGB}{232,102,102}
\definecolor{lightred}{RGB}{238,131,131}
\definecolor{bertiniblue}{RGB}{101,147,245}
\definecolor{bertinigreyblue}{RGB}{101,147,245}
\definecolor{bertinigreyred}{RGB}{232,102,102}
\definecolor{bertinivioletc}{RGB}{45,130,60}
\definecolor{bertinigreen}{RGB}{166,218,149}
\definecolor{bertiniorange}{RGB}{255, 116, 23}
\definecolor{OliveGreen}{RGB}{85,107,47}
\definecolor{NavyBlue}{RGB}{0,0,128}
\definecolor{bertiniviolet}{RGB}{210,145,178}
\definecolor{bertinigrey1}{RGB}{98,98,98}
\definecolor{bertinigrey2}{RGB}{211,211,211}
\definecolor{bertinigrey3}{RGB}{192,192,192}
\definecolor{bertinigrey4}{RGB}{169,169,169}

\newcommandx{\tikzdiagup}{
	\tikz {\draw[thick] (0,0)--(0.15,0.15); \draw (0,0) rectangle (0.15,0.15);}
}

\newcommandx{\gatecross}[1][1=0.5]{
	\pgfmathparse{#1/2.0}
	\let\x\pgfmathresult
	\draw[thick] (-\x,-\x) -- (\x,\x);
	\draw[thick] (\x,-\x) -- (-\x,\x);
}
\newcommandx{\gatesqu}[2][1=0.25,2=]{
	\pgfmathparse{#1/2.0}
	\let\x\pgfmathresult
	\ifthenelse{\equal{#2}{}}{
		\draw[thick, fill=white, rounded corners=2pt] (-\x,\x) rectangle (\x,-\x);
	}{
		\draw[thick, fill=#2, rounded corners=2pt] (-\x,\x) rectangle (\x,-\x);
	}
}

\newcommandx{\gatemark}[2][1=0.075,2=tr]{
	\pgfmathparse{#1}
	\let\l\pgfmathresult
	\ifthenelse{\equal{#2}{topleft}}{
		\draw[thick] (0,\l) -- ++(-\l,0) --++ (0,-\l);
	}{}
	\ifthenelse{\equal{#2}{topright}}{
		\draw[thick] (0,\l) -- ++(\l,0) --++ (0,-\l);
	}{}
	
	\ifthenelse{\equal{#2}{bottomleft}}{
		\draw[thick] (0,-\l) -- ++(-\l,0) --++ (0,\l);
	}{}
	\ifthenelse{\equal{#2}{bottomright}}{
		\draw[thick] (0,-\l) -- ++(\l,0) --++ (0,\l);
	}{}
	
}

\newcommandx{\roundgate}[6][1=0,2=0,3=1,4=topright,5=white,6=-1]{
	\pgfmathparse{#3}
	\let\l\pgfmathresult
	\begin{scope}[shift={(#1,#2)}]
		\gatecross[\l]
			\pgfmathparse{\l/2.0}
		\let\s\pgfmathresult
		\gatesqu[\s][#5]
		\pgfmathparse{\l*0.15}
		\let\m\pgfmathresult
	\ifthenelse{\equal{#6}{-1}}{		\gatemark[\m][#4]
	}{	\node at ({0},{0}) {\scalebox{1.3}{$#6$}};}
\end{scope}
}
\newcommandx{\wcirc}[2]{\begin{scope}
		\draw[fill=white] (#1,#2) circle (0.15);	\end{scope}} 
\newcommandx{\wcircc}[2]{\begin{scope}
		\draw[fill=white] (#1,#2) circle (0.13);	\end{scope}} 

\newcommandx{\wsqr}[2]{\begin{scope}
		\draw[fill=white,shift={(#1,#2)}] (-.13,.13) rectangle (.13,-.13);	\end{scope}} 
\newcommandx{\wsqrr}[2]{\begin{scope}
		\draw[fill=white,shift={(#1,#2)}] (-.11,.11) rectangle (.11,-.11);	\end{scope}} 
\newcommandx{\bcirc}[2]{\begin{scope}
		\draw[fill=black] (#1,#2) circle (0.15);	\end{scope}} 
\newcommandx{\thetastate}[4][1=0,2=0,3=1,4=]{
	\pgfmathparse{#3/2}
	\let\l\pgfmathresult
	\pgfmathparse{\l*0.15}
	\let\m\pgfmathresult
	\begin{scope}[shift={(#1,#2)}]
		\draw[thick] (0,0)--(\l,\l);
		\draw[thick] (0,0)--(-\l,\l);
		\ifthenelse{\equal{#4}{}}{
			\draw[fill=white] (0,0) circle (0.15);
		}{
			\draw[thick, fill=#4] (0,0) circle (0.15);
		}
	\end{scope}
}

\newcommandx{\thetastateflipped}[4][1=0,2=0,3=1,4=]{
	\pgfmathparse{#3/2}
	\let\l\pgfmathresult
	\pgfmathparse{\l*0.15}
	\let\m\pgfmathresult
	\begin{scope}[shift={(#1,#2)}]
		\draw[thick] (0,0)--(\l,-\l);
		\draw[thick] (0,0)--(-\l,-\l);
		\ifthenelse{\equal{#4}{}}{
			\draw[fill=white] (0,0) circle (0.15);
		}{
			\draw[thick, fill=#4] (0,0) circle (0.15);
		}
	\end{scope}
}

\newcommandx{\vertgate}[5][1=0,2=0,3=4,4=bertiniorange,5=topright]
{
	\begin{scope}[shift={(#1,#2)}]
		\ifthenelse{\equal{#3}{1}}{
			\roundgate[0][0][1][#5][#4]
		}{
			\foreach \n[evaluate=\n as \y using {2*\n-2}] in {1,...,#3}{
				\roundgate[0][\y][1][#5][#4]
			}
		}
	\end{scope}
}

\newcommandx{\tsfmatV}[8][1=0,2=0,3=l,4=4,5=tr,6=init,7=bertiniorange,8=topright]{
	\begin{scope}[shift={(#1,#2)}]
		\ifthenelse{\equal{#3}{l}}{
			\pgfmathsetmacro{\flag}{0}
		}{
			\pgfmathsetmacro{\flag}{1}
		}
		
		\foreach \y[evaluate=\y as \x using {mod(\y+\flag,2)}] in {1,...,#4}{
			\roundgate[\x][\y][1][#8][#7]
		}
		\ifthenelse{\equal{#5}{tr}}{
			\foreach \y[evaluate=\y as \x using {mod(\y+\flag,2)}] in {#4}{
				\draw [fill=white] (\x-0.5,\y+0.5) circle (0.15);
				\draw [fill=white] (\x+0.5,\y+0.5) circle (0.15);
			}
		}{}
		\ifthenelse{\equal{#6}{init}}{
			\thetastate[\flag][0][1][#7]
		}{}
	\end{scope}
}

\newcommandx{\leftriangle}[5][1=0,2=0,3=4,4=bertiniorange,5=topright]{
	\begin{scope}[shift={(#1,#2)}]
		\pgfmathsetmacro{\t}{#3}
		\pgfmathsetmacro{\steps}{ceil(\t/2)}
		\foreach \i[evaluate=\i as \x using -\t+2*\i-1, evaluate=\i as \ylim using \t-2*\i+2] in {1,...,\steps}{
			\foreach \y[evaluate=\y as \thisx using {\x+\y-1}] in {1,...,\ylim}{
				\roundgate[\thisx][\y][1][#5][#4]
			}
		}
	\end{scope}
}

\newcommandx{\rightriangle}[5][1=0,2=0,3=4,4=bertiniorange,5=topright]{
	\begin{scope}[shift={(#1,#2)}]
		\pgfmathsetmacro{\t}{#3}
		\pgfmathsetmacro{\steps}{ceil(\t/2)}
		\foreach \i[evaluate=\i as \x using -\t+2*\i-1, evaluate=\i as \ylim using \t-2*\i+2] in {1,...,\steps}{
			\foreach \y[evaluate=\y as \thisx using {-\x-\y+1}] in {1,...,\ylim}{
				\roundgate[\thisx][\y][1][#5][#4]
			}
		}
	\end{scope}
}

\newcommandx{\eigenVL}[8][1=0,2=0,3=l,4=5,5=tr,6=init,7=bertiniorange,8=topright]{
	\begin{scope}[shift={(#1,#2)}]
		\pgfmathsetmacro{\t}{#4}
		\leftriangle[0][0][\t][#7][#8]
		
		\ifthenelse{\equal{#6}{init}}{
			\drawinitstate[0][0][l][\t][#7]
		}{}
		
		\ifthenelse{\equal{#5}{tr}}{
			\draw[fill=white] \foreach \x in {0,...,\t} {(\x-0.5-\t,0.5+\x) circle (0.15)};
			\ifthenelse{\equal{#3}{r}}{
				\draw[fill=white] (0.5,\t+0.5) circle (0.15);
			}{}
		}{}
		\ifthenelse{\equal{#5}{parttr}}{
			\draw[fill=white] \foreach \x in {0,...,\t} {(\x-0.5-\t,0.5+\x) circle (0.15)};
		}{}
	\end{scope}
}

\newcommandx{\eigenVR}[8][1=0,2=0,3=l,4=5,5=tr,6=init,7=bertiniorange,8=topright]{
	\begin{scope}[shift={(#1,#2)}]
		\pgfmathsetmacro{\t}{#4}
		\rightriangle[0][0][\t][#7][#8]
		
		\ifthenelse{\equal{#6}{init}}{
			\drawinitstate[0][0][r][\t][#7]
		}{}
		
		\ifthenelse{\equal{#5}{tr}}{
			\draw[fill=white] \foreach \x in {0,...,\t}{(-\x+0.5+\t,0.5+\x) circle (0.15)};
			\ifthenelse{\equal{#3}{l}}{
				\draw[fill=white] (-0.5,\t+0.5) circle (0.15);
			}{}
		}{}
	\end{scope}
}

\newcommandx{\tra}[2][1]{\underset{#1}{\text{tr}}\left[#2\right]}

\newcommandx{\tsfmatDgate}[7][1=0,2=0,3=l,4=4,5=tr,6=bertiniorange,7=topright]
{
	\begin{scope}[shift={(#1,#2)}]
		\ifthenelse{\equal{#3}{l}}{
			\pgfmathsetmacro{\flag}{-1}
		}{
			\pgfmathsetmacro{\flag}{1}
		}
		\pgfmathsetmacro{\t}{#4}
		\foreach \i[evaluate=\i as \x using {\flag*\i}, evaluate=\i as \y using \i] in {1,...,\t}{
			\roundgate[\x][\y][1][#7][#6]
		}
		
		\ifthenelse{\equal{#5}{tr}}{
			\foreach \i[evaluate=\i as \x using {\flag*\i}, evaluate=\i as \y using \i] in {\t}{
				\draw [fill=white] (\x-0.5,\y+0.5) circle (0.15);
				\draw [fill=white] (\x+0.5,\y+0.5) circle (0.15);
			}  
		}{}
	\end{scope}
	
}

\newcommandx{\tsfmatD}[8][1=0,2=0,3=l,4=4,5=tr,6=init,7=bertiniorange,8=topright]{
	\begin{scope}[shift={(#1,#2)}]
		\ifthenelse{\equal{#6}{init}}{
			\thetastate[0][0][1][#7]
		}{}
		
		\ifthenelse{\equal{#3}{l}}{
			\pgfmathsetmacro{\flag}{-1}
		}{
			\pgfmathsetmacro{\flag}{1}
		}
		
		\pgfmathsetmacro{\t}{#4}
		\foreach \i[evaluate=\i as \x using {\flag*\i}, evaluate=\i as \y using \i] in {1,...,\t}{
			\roundgate[\x][\y][1][#8][#7]
		}
		
		\ifthenelse{\equal{#5}{tr}}{
			\foreach \i[evaluate=\i as \x using {\flag*\i}, evaluate=\i as \y using \i] in {\t}{
				\draw [fill=white] (\x-0.5,\y+0.5) circle (0.15);
				\draw [fill=white] (\x+0.5,\y+0.5) circle (0.15);
			}  
		}
		\ifthenelse{\equal{#5}{parttr}}{
			\foreach \i[evaluate=\i as \x using {\flag*\i}, evaluate=\i as \y using \i] in {\t}{
				\draw [fill=white] (\x+0.5*\flag,\y+0.5) circle (0.15);
			}  
		}
		{}
	\end{scope}
}

\newcommandx{\drawinitstate}[5][1=0,2=0,3=l,4=4,5=bertiniorange]{
	\pgfmathsetmacro{\t}{#4}
	\begin{scope}[shift={(#1,#2)}]
		\pgfmathsetmacro{\steps}{ceil((\t-1)/2)}
		\ifthenelse{\equal{#3}{l}}{
			\foreach \i[evaluate=\i as \x using -\t+2*\i] in {0,...,\steps}{
				\thetastate[\x][0][1][#5]
			}
		}{
			\foreach \i[evaluate=\i as \x using -\t+2*\i] in {0,...,\steps}{      
				\thetastate[-\x][0][1][#5]
			}
		}
	\end{scope}
}

\newcommandx{\drawinitstateflipped}[5][1=0,2=0,3=l,4=4,5=bertiniorange]{
	\pgfmathsetmacro{\t}{#4}
	\begin{scope}[shift={(#1,#2)}]
		\pgfmathsetmacro{\steps}{ceil((\t-1)/2)}
		\ifthenelse{\equal{#3}{l}}{
			\foreach \i[evaluate=\i as \x using -\t+2*\i] in {0,...,\steps}{
				\thetastateflipped[\x][0][1][#5]
			}
		}{
			\foreach \i[evaluate=\i as \x using -\t+2*\i] in {0,...,\steps}{      
				\thetastateflipped[-\x][0][1][#5]
			}
		}
	\end{scope}
}

\newcommandx{\eigenDL}[6][1=0,2=0,3=l,4=4,5=bertiniorange,6=topright]{
	\begin{scope}[shift={(#1,#2)}]
		\pgfmathsetmacro{\t}{#4}
		\ifthenelse{\equal{#3}{l}}{
			\eigenVL[0][0][l][\t][tr][init][#5][#6]
			\pgfmathsetmacro{\t}{#4-1}
			\rightriangle[1][0][\t][#5][#6]
			\drawinitstate[1][0][r][\t][#5]
		}{
			\begin{scope}[shift={(-0.5,0.5)}]
				\foreach \i[evaluate=\i as \x using \i, evaluate=\i as \y using \i] in {0,...,\t}{      
					\draw (\x,\y)--++(0.5,0);
					\draw[fill=white] (\x,\y) circle (0.15);
				}
			\end{scope}
		}
	\end{scope}
}

\newcommandx{\eigenDR}[6][1=0,2=0,3=l,4=4,5=bertiniorange,6=topright]{
	\begin{scope}[shift={(#1,#2)}]
		\pgfmathsetmacro{\t}{#4}
		\ifthenelse{\equal{#3}{r}}{
			\eigenVR[0][0][r][\t][tr][init][#5][#6]
			\pgfmathsetmacro{\t}{#4-1}
			\leftriangle[-1][0][\t][#5][#6]
			\drawinitstate[-1][0][l][\t][#5]
		}{
			\begin{scope}[shift={(0.5,0.5)}]
				\foreach \i[evaluate=\i as \x using \i, evaluate=\i as \y using \t-\i] in {0,...,\t}{      
					\draw (\x,\y)--++(0.5,0);
					\draw[fill=white] (\x+0.5,\y) circle (0.15);
				}
			\end{scope}
		}
	\end{scope}
}

\newcommandx{\idonpurity}[2][1=0,2=0]
{
	\begin{scope}[shift={(#1,#2)}]
		\draw[thick] (-0.5,0)--++(-0.1,0.1)--++(0,0.2)--++(0.1,-0.1);
		\draw[thick] (-0.5,0.4)--++(-0.1,0.1)--++(0,0.2)--++(0.1,-0.1);
		\draw[thick] (0.5,0)--++(0.1,0.1)--++(0,0.2)--++(-0.1,-0.1);
		\draw[thick] (0.5,0.4)--++(0.1,0.1)--++(0,0.2)--++(-0.1,-0.1);
	\end{scope}
}

\newcommandx{\swaponpurity}[2][1=0,2=0]
{
	\begin{scope}[shift={(#1,#2)}]
		\draw[thick] (-0.5,0)--++(-0.2,0.2)--++(0,0.6)--++(0.2,-0.2);
		\draw[thick] (-0.5,0.2)--++(-0.075,0.075)--++(0,0.2)--++(0.075,-0.075);
		\draw[thick] (+0.5,0)--++(+0.2,0.2)--++(0,0.6)--++(-0.2,-0.2);
		\draw[thick] (+0.5,0.2)--++(+0.075,0.075)--++(0,0.2)--++(-0.075,-0.075);
	\end{scope}
}

\newcommandx{\hook}[4][1=0,2=0,3=t,4=l]{
	\begin{scope}[shift={(#1,#2)}]
		\ifthenelse{\equal{#3}{t}}{
			\ifthenelse{\equal{#4}{l}}{\draw[thick] (0.5,-0.5) arc (45:-90:0.15);}{\draw[thick] (0.5,-0.5) arc (45:270:0.15);}
		}{\ifthenelse{\equal{#4}{l}}{\draw[ thick] (0.5,-0.5) arc (-45:90:0.15);}{\draw[ thick] (0.5,-0.5) arc (315:90:0.15);}
		}
	\end{scope}
}

\newcommandx{\hhook}[4][1=0,2=0,3=t,4=l]{
	\begin{scope}[shift={(#1,#2)}]
		\ifthenelse{\equal{#3}{t}}{
			\ifthenelse{\equal{#4}{l}}{\draw[thick] (0.5,-0.5) arc (-45:175:0.15);}{\draw[thick] (0.5,-0.5) arc (225:0:0.15);}
		}{\ifthenelse{\equal{#4}{l}}{\draw[ thick] (0.5,-0.5) arc (-45:180:-0.15);}{\draw[ thick] (0.5,-0.5) arc (45:-180:0.15);}
		}
	\end{scope}
}

\newcommandx{\Pproj}[3][3=$P_\Lambda$]{
\begin{scope}[shift={(#1-.5,#2-1)}]
\draw[thick,fill=white] (0,0)rectangle (1,2);
\draw[thick] (0,1.5)--(-.5,1.5);
\draw[thick] (1,1.5)--(1.5,1.5);
\draw[thick] (0,.5)--(-.5,.5);
\draw[thick] (1,.5)--(1.5,.5);
\node[scale=2] at (.5,1) {#3};
\end{scope}}

\definecolor{FcolU}{rgb}{0.71,0.78,0.91}
\definecolor{colLines}{rgb}{0.31,0.31,0.31}
\definecolor{colVMPSLines}{rgb}{0.11,0.11,0.11}
\definecolor{IcolUc}{rgb}{0.71,0.41,0.42}
\definecolor{IcolU}{rgb}{0.71,0.8,0.76}
\definecolor{IcolVMPSc}{rgb}{0.73,0.69,0.7}
\definecolor{IcolVMPS}{rgb}{0.81,0.77,0.78}
\definecolor{colObs}{rgb}{1.,1.,1.}

\renewcommand{\comment}[1]{\ifdef{0}{}{#1}}

\newcommandx{\eightlegs}[2][1=0,2=0]{
	\begin{scope}[shift={(#1,#2)}]
		\foreach \x in {1,...,8}{
			\draw (\x, 0)--++(0,0.25);
			\draw[fill] (\x,0) circle (0.05);
		}
		\foreach \x in {1,3}{
			\pgfmathsetmacro\result{2*\x-1} 
			\node () at (\result,-0.5) {$i_{\x}$};
			\pgfmathsetmacro\result{2*\x}
			\node () at (\result,-0.5) {$j_{\x}$};	
		}
		\foreach \x in {2,4}{
			\pgfmathsetmacro\result{2*\x} 
			\node () at (\result,-0.5) {$i_{\x}$};
			\pgfmathsetmacro\result{2*\x-1}
			\node () at (\result,-0.5) {$j_{\x}$};	
		}
	\end{scope}
}

\newcommandx{\MPSinitialstate}[5][1=0,2=0,3=bertiniorange,4=topright,5=-1]{
\begin{scope}[shift={(#1,#2)},rounded corners=1.5pt]
	\draw[black,thick,fill=#3] 
	(-0.25,.25)--++(.5,0)--++(0,-.3)--++(-.5,0)--cycle;
	\draw[thick] (-.25,.25)--++(-.25,.25);
	\draw[thick] (.25,.25)--++(.25,.25);
	\draw[very thick] (-1,.-.05)--++(2,0);
\ifthenelse{\equal{#5}{-1}}{
	\ifthenelse{\equal{#4}{topright}}{\draw[thick,rounded corners=0.3]
	(-.1,.15)--++(.2,0)--++(0,-0.1); }{}
	\ifthenelse{\equal{#4}{topleft}}{\draw[thick,rounded corners=0.3]
	(.1,.15)--++(-.2,0)--++(0,-0.1); }{}
	\ifthenelse{\equal{#4}{bottomleft}}{\draw[thick,rounded corners=0.3]
	(.1,-.15)--++(-.2,0)--++(0,0.1); }{}	\ifthenelse{\equal{#4}{bottomright}}{\draw[thick,rounded corners=0.3]
	(-.1,-.15)--++(.2,0)--++(0,0.1); }{}}{			\node at ({0},{0.085}) {\scalebox{1.}{{$#5$}}};}
\end{scope}
}

\newcommandx{\Cmatrix}[6][1=0,2=0,3=2,4=bertiniorange,5=,6=topright]{
	\pgfmathsetmacro\result{#3-1} 
	\begin{scope}[shift={(#1,#2)}]
		\foreach \i in {0,...,\result}
		{\foreach \j in {0,...,\i}
			{\roundgate[\i+\j][\i-\j][1][#6][#4]}
		}
		\ifthenelse{\equal{#5}{init}}{
			\foreach \i in {0,...,#3}
			{
				\MPSinitialstate[-1+2*\i][-1][#4]
			}
		}{}
	\end{scope}
}

\newcommand{\mcirc}{\fineq[-0.8ex][0.7][1]{\cstate[0][0][]}}
\renewcommand{\bcirc}{\fineq[-0.8ex][0.7][1]{\cstate[0][0][][black]}}

\newcommandx{\cstate}[4][1=0,2=0,3= ,4=white]{
	\begin{scope}[shift={(0,0)}]
				\draw[fill=#4,thick] (#1,#2) circle (0.13);
				\node[scale=1.1] at (#1,#2) {$#3$};
\end{scope}
}
\newcommandx{\sqrstate}[4][1=0,2=0,3= ,4=white]{
	\begin{scope}[shift={(#1,#2)}]
		\draw[thick,fill=#4] (-0.13,-0.13) rectangle (0.13,0.13) ;
		\node[scale=1.1] at (0,0) {$#3$};
	\end{scope}
}
\newcommandx{\pairproduct}[2][1=0,2=0]
{\begin{scope}[shift={(#1 ,#2)}]
\draw[thick] (-.5,.5) arc(-135:-45:1/1.414);
\sqrstate[0][.5-.1414][][black]	
\end{scope}
}
\newcommandx{\bellpair}[2][1=0,2=0]
{\begin{scope}[shift={(#1 ,#2)}]
		\draw[thick] (-.5,.5) arc(-135:-45:1/1.414);
	\end{scope}
}
\newcommandx{\charge}[3][1=0,2=0,3=black]
{
	\ifthenelse{\equal{#3}{blue}}{\def \chargecolor{bertinigreyblue}
	}{
	\ifthenelse{\equal{#3}{red}}{\def \chargecolor{bertinigreyred}}{\def \chargecolor {#3}}}
\begin{scope}[shift={(#1 ,#2)}]
	\draw[ fill=\chargecolor] circle (0.08);        
\end{scope}
}
\newcommandx{\trianglediag}[7][1=0,2=0,3=1,4=bertiniorange,5= ,6=-1,7=topright]
{\begin{scope}[shift={(#1 ,#2)}]
	\foreach \i in {0,...,#3}
	{	\foreach \j in {0,...,\i}
		{	\roundgate[-\j+2*\i][\j][1][topright][#4][#6]
		}
	}
	\foreach \i in {-1,...,#3}
	{\ifthenelse{\equal{#5}{bellpair}}{\bellpair[\i*2+1][-1]}{\ifthenelse{\equal{#5}{pairproduct}}{\pairproduct[\i*2+1][-1]}{\MPSinitialstate[\i*2+1][-1][#4][#7][#6]}}}
\end{scope}
}
\newcommandx{\projectorleg}[4][1=0,2=0,3=R,4=left]
{
	\begin{scope}[shift={(#1 ,#2)}]
		{\ifthenelse{\equal{#3}{R}}{		\draw[thick] (-.25,-.25)--++(.5,.5);}{\ifthenelse{\equal{#3}{L}}{		\draw[thick] (-.25,.25)--++(.5,-.5);}{\draw[thick] (-.25,0)--++(.5,0);}}
		}
	\draw[thick, fill=white] circle (0.13);
	\ifthenelse{\equal{#4}{right}}{
		\draw[thick] (.0,.07)--++(.07,0)--++(0,-.07);
		\node[scale=0.5] at (0,-.02) {$\alpha$};
	}{}
	\ifthenelse{\equal{#4}{left}}{	
	\draw[thick] (0,.07)--++(-.07,0)--++(0,-.07);
	\node[scale=0.5] at (0,-.03) {$\beta$};
	}{}
	\end{scope}
}
\usetikzlibrary{patterns,decorations.pathreplacing}

\newcommand{\lt}[0]{{\textsc{l}}}
\newcommand{\rt}[0]{{\textsc{r}}}

\pgfplotsset{
	colormap={lightbluepalette}{ rgb255=(196, 167, 175) rgb255=(135, 206, 250) rgb255=(115, 194, 251) rgb255=(124,158,217) rgb255=(96, 130, 182)}
}
\pgfplotsset{
    colormap={springpastels}{ rgb255=(253, 127, 111) rgb255=(126, 176, 213) rgb255=(178, 224, 97) rgb255=(189, 126, 190) rgb255=(255, 181, 90)  rgb255=(190, 185, 219) rgb255=(253, 204, 229) rgb255=(139, 211, 199)}
    }
\definecolor{pastelred}{RGB}{253, 127, 111}
\definecolor{pastelblue}{RGB}{126, 176, 213}
\definecolor{pastelgreen}{RGB}{178, 224, 97}
\definecolor{trollgreen}{RGB}{200, 253, 97}
\definecolor{newcol}{RGB}{189, 126, 190}